\documentclass[10pt,a4paper]{article}

\usepackage[T1]{fontenc}
\usepackage{lmodern}        
\usepackage{geometry}       
\usepackage{setspace}
\usepackage{amsmath,amssymb,amsthm}
\usepackage{graphicx}
\usepackage{hyperref}
\usepackage[numbers]{natbib}
\usepackage{amssymb,amsmath,bbm,ragged2e,graphicx,physics,enumerate,color,multirow,booktabs,bm,titlesec,subcaption,natbib}
\usepackage[export]{adjustbox}

\newcommand{\bathh}{\textrm{h}}
\newcommand{\bathw}{\textrm{w}}
\newcommand{\bathc}{\textrm{c}}

\newcommand{\ketg}{\ket{\textrm{g}}}
\newcommand{\keth}{\ket{\textrm{h}}}
\newcommand{\ketc}{\ket{\textrm{c}}}
\newcommand{\brag}{\bra{\textrm{g}}}
\newcommand{\brah}{\bra{\textrm{h}}}
\newcommand{\brac}{\bra{\textrm{c}}}

\title{Non-equilibrium Dynamics of Three-Level Absorption Refrigerator at Third-Order Liouvillian Exceptional Points}
\author{Jingyi Gao\\
Department of Physics, the University of Tokyo\\5-1-5 Kashiwanoha, Kashiwa, Chiba, 277-8574, Japan.\\
Naomichi Hatano\\
Institute of Industrial Science, the University of Tokyo\\5-1-5 Kashiwanoha, Kashiwa, Chiba, 277-8574, Japan}
\date{\today}

\begin{document}
\maketitle

\begin{abstract}
We analyze the influence of Liouvillian exceptional points (LEPs) in the three-level quantum absorption refrigerator, putting emphasis on the non-equilibrium process before the convergence to the steady state. We search for the second-order and third-order LEPs in the system with two types of couplings. Focusing on the third-order LEPs, we analyze the damping of the system state in the long term analytically and numerically. In addition, we analyze the damping of heat currents and the influence of the non-equilibrium process in the heat extraction from the cold bath. Critical damping at LEPs of both the system state and the heat currents is achieved, implying the fastest convergence to the equilibrium system. During the non-equilibrium process, we find that much heat transfer from the cold bath to the hot bath with less energy cost of the work bath is achieved at the third-order LEP, leading to better performance of the refrigerator.
\end{abstract}

\maketitle
\section{Introduction}
\label{Intro}

The theory of open quantum systems has been attracting much attention recently~\cite{schaller2014open, 9780199213900, 978-3-642-23354-8, gardiner2004quantum, PhysRevB.87.201402, moiseyev2011non, PhysRevB.56.8651, rotter2009non, 10.1063/1.4904200}. 
By using various approximations and conditions, we can derive several master equations for describing the time evolution of open quantum systems~\cite{PRXQuantum.5.020202, carmichael2009open, PhysRevA.75.022103, dann2018time, 10.1143/PTP.20.948, 10.1063/1.1731409, BF01040100, doi:10.1142/S1230161222500044, PhysRevResearch.3.013165}. A famous one is the Gorini-Kossakowski-Sudarshan-Lindblad (GKSL) master equation, which is under the Born-Markov approximation with weak couplings between the system and the environment and prompt equilibration of the environment~\cite{1.522979, BF01608499, 10.1063/1.5115323, doi:10.1142/S1230161217400017, Hofer_2017}. In the GKSL master equation, the information about the evolution, including the Hamiltonian and the dissipation, can be described by the Liouvillian superoperator.  
Although the system Hamiltonian is Hermitian, because of the dissipation to the environment, the Liouvillian superoperator can be non-Hermitian. 

In the non-Hermitian system, exceptional points (EPs) can exist. Different from the diabolic points (DPs) in the Hermitian system, in which the eigenvalues are degenerate with their eigenvectors orthogonal to each other, in the non-Hermitian system at EPs, the eigenvalues are degenerate with one coalescent eigenvector~\cite{heiss2004exceptional, berry2004physics, RevModPhys.93.015005, doi:10.1126/science.aar7709, Heiss_2012, PhysRevLett.103.134101, PhysRevB.109.085311, PhysRevLett.127.107402, PhysRevA.110.012226}. The EP of the Liouvillian superoperator is referred to as the Liouvillian exceptional point (LEP). Since LEP can exist even in systems with Hermitian Hamiltonians, it has become a hot topic recently~\cite{PRXQuantum.2.040346, Hatano18082019, PhysRevA.100.062131}.

As an important application of open quantum systems, quantum thermal machines have also attracted much attention in recent years since they can show significant advantages compared to classical thermal machines~\cite{Bhattacharjee2021, PRXQuantum.2.030310, PhysRevB.102.155407, PhysRevLett.2.262, kosloff1984quantum, RAlicki_1979, geusic1967quantum, geva1996quantum, partovi1989quantum, kosloff2000quantum, palao2001quantum, lloyd1997quantum, kieu2004second, PhysRevE.87.012140, quan2007quantum, PhysRevResearch.5.023066, PhysRevResearch.6.023172}. There are various categories of quantum thermal machines. 
Quantum absorption refrigerator~\cite{Maslennikov2019, PhysRevLett.108.070604, PhysRevB.102.235427, PhysRevB.104.075442, PhysRevE.87.042131, Correa2014, PhysRevE.92.062101} is a type of quantum cooler, which only comprises several heat baths and an internal system without work components. Without considering the process of work production, the analysis of the absorption refrigerator is easier than other quantum thermal machines. In addition, similarly to the quantum amplifier~\cite{stenholm1986theory, haus1962quantum, caves1982quantum, loudon1984properties}, the quantum absorption refrigerator is not based on a thermal cycle, which also simplifies the operation protocol.

However, hitherto, the research on non-equilibrium quantum thermal machines lacks discussions on their non-Hermitian dynamics towards the steady state. Although some previous works studied non-Markovian quantum thermal machines with simple structures~\cite{PhysRevResearch.3.023078, PhysRevResearch.5.023066}, research on non-Hermitian quantum thermal machines is rare~\cite{santos2023pt, zhang2022dynamical, PRXQuantum.2.040346}. 

Besides, since the quantum absorption refrigerator is a relatively new topic, previous research mostly focused on the equilibrium system after achieving the steady state. However, when there are interactions between the internal system and the external environment, there must be a non-equilibrium stage before achieving the steady state. With a long duration or strong influence of the non-equilibrium process, the non-Hermitian evolution before the system converges to the steady state is not negligible. Although it is important in practice, the non-equilibrium process before achieving the equilibrium state lacks discussion. 

In the present work, we consider the non-equilibrium process of the three-level quantum absorption refrigerators~\cite{PhysRevE.97.052145, PhysRevE.100.062112, PhysRevE.101.062121, PhysRevE.92.012136, Correa2014}.
Since there are only system-bath interactions between the internal system and the external environment, if we consider the Markovian conditions for the system-bath interactions, the Liouvillian superoperator derived from the GKSL master equation plays a significant role in such a quantum absorption refrigerator. Since LEPs represent special conditions of the Liouvillian superoperator and show unique properties in the non-equilibrium process, it is worth exploring the quantum absorption refrigerators at LEPs.

First, we search for the second- and third-order LEPs in the one-coupling three-level quantum absorption refrigerators. Then we analyze the influences of the LEPs in the quantum absorption refrigerators from various perspectives, including the damping of the system state and heat current, the non-equilibrium process before convergence to the steady state, and the performance of the refrigerator impacted by the non-equilibrium process. Although we assume that the Hamiltonian of the system itself is still Hermitian and can be analyzed by the GKSL master equation, due to the unique properties of the LEP, critical damping and better performances can be achieved in the non-equilibrium evolution from the initial state to the steady state.

The paper is organized as follows. In Secs.~\ref{Sec1}--\ref{Sec3}, we define the models of the quantum absorption refrigerator based on a qutrit and three heat baths and search for the third-order LEP of the systems. 
In Sec.~\ref{Sec1}, we describe the operation protocol of the three-level quantum absorption refrigerator and construct the Liouvillian superoperator from the GKSL master equation. In Sec.~\ref{Sec2}, we derive the conditions of the third-order LEPs. In Sec.~\ref{Sec3}, we describe the definitions of heat current and the coefficient of performance (COP) in the quantum absorption refrigerator.
In Sec.~\ref{Sec4}, we confirm the critical damping of the system state and heat current at LEPs. 
In Sec.~\ref{Sec5}, we analyze the performance of the refrigerator at the third-order LEP compared to the steady state and demonstrate that the non-equilibrium process can improve the performance. Finally, we conclude in Sec.\ref{Conclusion}.

\section{Model and Dynamics}
\label{Sec1}

\begin{figure*}
	\centering
	\begin{subfigure}{0.38\textwidth}
		\includegraphics[width=0.8\textwidth]{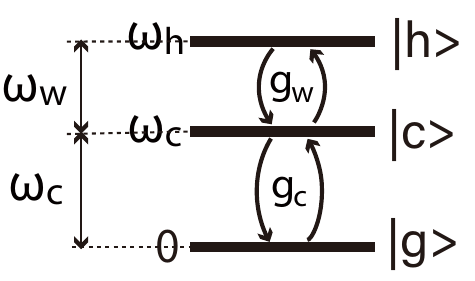}
		\caption{ }
		\label{energylevels}
	\end{subfigure}
	\begin{subfigure}{0.58\textwidth}
		\includegraphics[width=0.8\textwidth]{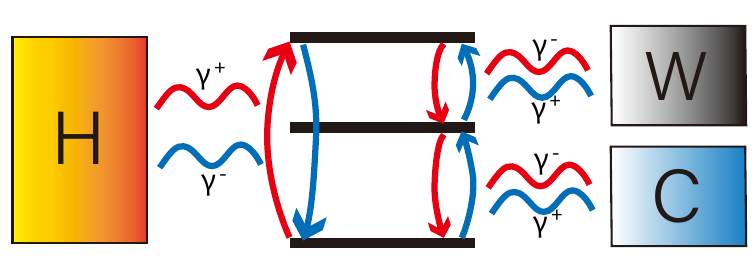}
		\caption{ }
		\label{interaction}
	\end{subfigure}
	\caption{Schematic views of the three-level absorption refrigerator. (a)~Energy levels and couplings of the internal three-level system. (b)~Interaction between the internal three-level system and the external three heat baths.}
	\label{ThreeLevelModel}
\end{figure*}

We construct our three-level quantum absorption refrigerator with a three-level internal system and three heat baths, as shown in Fig.~\ref{ThreeLevelModel}. 
Considering the possible couplings between the neighboring energy levels, we define the Hamiltonians of the internal system, the heat baths, and the couplings between them, respectively, as follows:
\begin{align}
\hat{H}_\textrm{s}&
=\mqty(
0 & g_\bathc  & 0 \\
g_\bathc & \omega_\bathc & g_\bathw\\
0  & g_\bathw & \omega_\bathh
),\\
\hat{H}_\textrm{bath}&=\sum_{k,\alpha=\bathh,\bathw,\bathc}\epsilon_{k\alpha}\hat{b}_{k\alpha}^\dagger\hat{b}_{k\alpha}, \\
\hat{H}_{\textrm{int}}&=\sum_{k,\alpha=\bathh,\bathc}V_{k \alpha}(\ketg \bra{\alpha}+\ket{\alpha} \brag)(\hat{b}_{k \alpha}+\hat{b}_{k \alpha}^\dagger )\nonumber\\
&+\sum_{k}V_{k \bathw}(\ketc \brah +\keth\brac)(\hat{b}_{k \bathw}+\hat{b}_{k \bathw}^\dagger ),
\end{align}
where the basis of the system Hamiltonian $\hat{H}_\textrm{s}$ is given by $\mqty(
\ketg &\ketc &\keth
)^T$.
The transfer amplitude within the internal system is $g_\bathc$ between the ground state $\ketg$ and the lower excited state $\ketc$ while $g_\bathw$ between the lower excited state $\ketc$ and the higher excited state $\keth$. We assume that all parameters in $\hat{H}_\textrm{s}$ are positive.
In all heat baths, $\hat{b}_{k \bathw}$ and $\hat{b}_{k \bathw}^\dagger$ are Boson creation and annhilation operators, respectively, and $\epsilon_{k\alpha}$ are their dispersion relation.
The hot bath $\alpha=\bathh$ influences the internal system by transferring the energy between the ground state $\ketg$ and the higher excited state $\keth$ with the strength $V_{k \bathh}$, the cold bath $\alpha=\bathc$ between the ground state $\ketg$ and the lower excited state $\ketc$ with $V_{k \bathc}$, and the work bath $\alpha=\bathw$ between the lower excited state $\ketc$ and the higher excited state $\keth$ with $V_{k \bathw}$. 

In the present model, we assume that the interaction between the internal system and the heat baths is weak and the equilibration of the baths is quick, so that the Born-Markov approximation is valid~\cite{9780199213900, 978-3-642-23354-8}. Therefore, the time evolution of the internal system follows the Gorini-Kossakowski-Sudarshan-Lindblad (GKSL) master equation~\cite{978-3-642-23354-8, 9780199213900}:
\begin{align}
\label{L}
\dot{\rho}(t)=\hat{\mathcal{L}}\rho(t)=-i\left[\hat{H},\rho(t)]+[\sum_{\alpha=\bathh,\bathc}(\gamma_\alpha^+\hat{\mathcal{D}}[\ketg \bra{\alpha}] + \gamma_\alpha^- \hat{\mathcal{D}}[\ket{\alpha} \brag])+
\gamma_\bathw^+\hat{\mathcal{D}}[\ketc \brah] + \gamma_\bathw^- \hat{\mathcal{D}}[\keth \brac\right]\rho(t),
\end{align}
where the basis of the density matrix is $\mqty(
\ketg &\ketc &\keth
)^T$ and
\begin{align}
\hat{\mathcal{D}}[\hat{o}]\rho=&\frac{1}{2}(2\hat{o}\rho\hat{o}^\dagger-\hat{o}^\dagger\hat{o}\rho-\rho\hat{o}^\dagger\hat{o}).
\end{align}
Note that all parameters $\gamma_\alpha^\pm$ must be positive so that the time evolution can be Markovian.

In what follows, we represent the $3\times 3$ density matrix as a nine-dimensional Hilbert-Schmidt vector \par
$\begin{bmatrix}
\rho_{\textrm{gg}} & \rho_{\textrm{gc}} &\rho_\textrm{{gh}} &\rho_{\textrm{cg}} &\rho_{\textrm{cc}} &\rho_{\textrm{ch}} &\rho_{\textrm{hg}} &\rho_{\textrm{hc}} &\rho_{\textrm{hh}} \end{bmatrix}^T$, and then represent the Liouvillian superoperator $\hat{\mathcal{L}}$ of the three-level quantum absorption refrigerator in the form of a $9\times 9 $ matrix:
\begin{align}
\hat{\mathcal{L}}=
\label{L9}
\begin{bmatrix}
-\gamma_\bathh^- -\gamma_\bathc^- & i g_\bathc &0 &-i g_\bathc &\gamma_\bathc^+ &0 &0 &0 &\gamma_\bathh^+ \\
i g_\bathc &i \omega_\bathc - \Gamma_\bathc &i g_\bathw &0 &-i g_\bathc &0 &0 &0 &0 \\
0 &i g_\bathw &i \omega_\bathh -\Gamma_\bathh &0 &0 &-i g_\bathc &0 &0 &0 \\
-i g_\bathc &0 &0 &-i\omega_\bathc-\Gamma_\bathc &i g_\bathc &0 &-i g_\bathw &0 &0 \\
\gamma_\bathc^- &-i g_\bathc &0 &i g_\bathc &-\gamma_\bathc^+-\gamma_\bathw^- &i g_\bathw &0 &-i g_\bathw &\gamma_\bathw^+ \\
0 &0 &-i g_\bathc &0&i g_\bathw &i\omega_\bathw-\Gamma_\bathw &0 &0 &-i g_\bathw \\
0 &0 &0 &-i g_\bathw &0 &0 &-i\omega_\bathh-\Gamma_\bathh&i g_\bathc &0\\
0 &0 &0 &0 &-i g_\bathw &0 &i g_\bathc &-i\omega_\bathw-\Gamma_\bathw &ig_\bathw\\
\gamma_\bathh^- &0 &0 &0 &\gamma_\bathw^- &-i g_\bathw &0 &i g_\bathw &-\gamma_\bathw^+ -\gamma_\bathh^+
\end{bmatrix},
\end{align}
where
\begin{align}
&\Gamma_\bathc=\frac{1}{2}(\gamma_\bathh^-+\gamma_\bathw^-+\gamma_\bathc^+ +\gamma_\bathc^-), \\
&\Gamma_\bathw=\frac{1}{2}(\gamma_\bathh^++\gamma_\bathw^++\gamma_\bathw^- +\gamma_\bathc^+), \\
&\Gamma_\bathh = \frac{1}{2}(\gamma_\bathh^- +\gamma_\bathw^+ +\gamma_\bathh^+ +\gamma_\bathc^-).
\end{align}
We thus represent the GKSL equation~(\ref{L}) as a $9\times 9$ matrix equation.

\begin{figure}
	\centering
	\begin{subfigure}{0.42\textwidth}
		\includegraphics[width=0.7\textwidth]{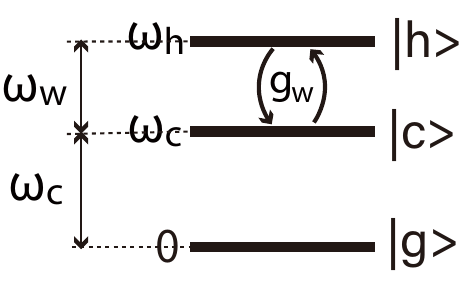}
		\caption{ }
		\label{energylevelgw}
	\end{subfigure}
	\begin{subfigure}{0.42\textwidth}
		\includegraphics[width=0.7\textwidth]{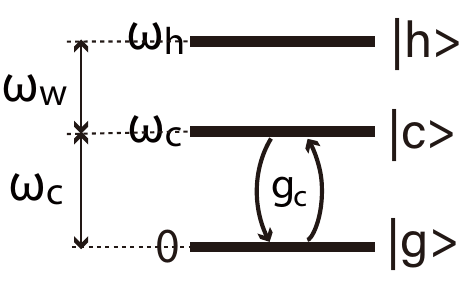}
		\caption{ }
		\label{energylevelgc}
	\end{subfigure}
	\caption{Schematic views of the three-level system with (a)~$g_\bathw$-coupling and (b)~$g_\bathc$-coupling.}
	\label{EnergyLevelgwgc}
\end{figure}

Since it is impossible to solve the general solution of a ninth-degree polynomial equation, we reduce the system with two assumptions to obtain general results in our model. First, we assume that the dissipation rates $\gamma^\pm$ are the same for every bath $\alpha$ in order to reduce the number of parameters:
\begin{align}
\gamma_\alpha^\pm = \gamma^\pm \text{ for }\alpha = \bathh,\bathc,\bathw.
\end{align}

Second, we make the system simpler by suppressing one of the two internal couplings, thereby reducing the Liouvillian superoperator $\hat{\mathcal{L}}_{9\times 9}$ to simpler forms. We consider two situations: 
(i) $g_\bathc=0$ and $g_\bathw \neq 0$, as in Fig.~\ref{energylevelgw}; (ii) $g_\bathw=0$ and $g_\bathc\neq 0$, as in Fig~\ref{energylevelgc}. In both cases, we can block-diagonalize the Liouvillian superoperator $\hat{\mathcal{L}}_{9\times 9}$ into a $5\times 5$ matrix $\hat{\mathcal{L}}_{5\times 5}$ and two $2\times 2$ matrices $\hat{\mathcal{L}}_{2\times 2}$ and $\hat{\mathcal{L}}^*_{2\times 2}$:
\begin{align}
\mathcal{L}_{9 \times 9} = \begin{bmatrix}
\mathcal{L}_{2 \times 2} &0 & 0\\
0 & \mathcal{L}_{2 \times 2}^* &0\\
0 &0 &\mathcal{L}_{5 \times 5}
\end{bmatrix}.
\end{align}
As detailed in Appendix~\ref{A1}, each system of (i) and (ii) can be reduced to an easily solvable system with four parameters $\{\omega_\alpha, g_\alpha, \gamma^+,\gamma^-\}$. We let $\alpha=\bathc$ or $\bathw$ denote the $g_\alpha$-coupling system.

\section{Eigensystem of the Liouvillian Block}
\label{Sec2}

As described in Appendix~\ref{A1}, the steady state with the zero-eigenvalue always appears in the $5\times 5$ block $\hat{\mathcal{L}}_{5\times 5}$ and the physical quantities influencing the heat current only exist in the block. Therefore, we focus our analysis on the $5\times 5$ matrix block $\hat{\mathcal{L}}_{5\times 5}$, ignoring the other two $2 \times 2$ matrices in the following sections. 

The distribution of the eigenvalues is given by the equation of the determinant $\mathrm{Det}|\hat{\mathcal{L}}_{5\times 5}-\lambda \mathbb{I}_5|=0$, where $\mathbb{I}_5$ denotes the $5\times 5$ identity matrix. For both $g_\bathw$- and $g_\bathc$-coupling systems, the determinant can be factorized into three terms as
\begin{align}
\label{L5Det}
\mathrm{Det}|\hat{\mathcal{L}}_{5\times 5}-\lambda \mathbb{I}|=-\lambda (\lambda -\lambda_1) F_3(\lambda),
\end{align}
which produc five eigenvalues, $0$, $\lambda$, and three roots of the cubic polynomial equation $F_3(\lambda)=0$.

\subsection{Distribution of Eiegnavlues}
\label{Sec2-1}

\begin{figure*}
	\centering
	\begin{subfigure}{0.24\textwidth}
		\includegraphics[width=1\textwidth]{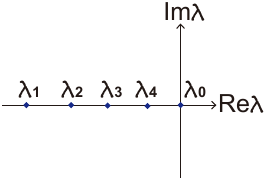}
		\caption{ }
		\label{nonEP1}
	\end{subfigure}
	\begin{subfigure}{0.24\textwidth}
		\includegraphics[width=1\textwidth]{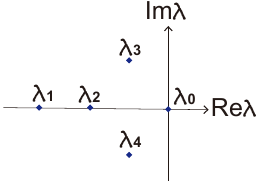}
		\caption{ }
		\label{nonEP2}
	\end{subfigure}
	\begin{subfigure}{0.24\textwidth}
		\includegraphics[width=1\textwidth]{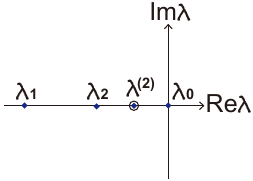}
		\caption{ }
		\label{EP2}
	\end{subfigure}
	\begin{subfigure}{0.24\textwidth}
		\includegraphics[width=1\textwidth]{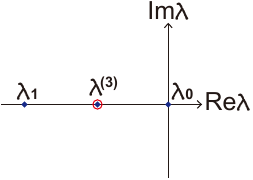}
		\caption{ }
		\label{EP3}
	\end{subfigure}
	\caption{Possible Distribution of the five eigenvalues with all real-positive parameters $\{\gamma^\pm, \omega_\alpha, g_\alpha\}$. (a) and (b) are possible distributions without LEPs, (c) is the distribution with a second-order LEP, and (d) is the distribution with a third-order LEP. Note that the sequence of the eigenvalue indices may be different. In (a), there are various orderings of the four real eigenvalues, and $\lambda_1$ is not always the smallest one. In (b) and (c), $\lambda_1$ and $\lambda_2$ may be greater than $\textrm{Re}[\lambda_{3,4}]$ and $\lambda^{(2)}$. In (d),  $\lambda_1$ may be greater than $\lambda_{(3)}$.}
	\label{Eigs}
\end{figure*}

The zero eigenvalue represents the steady state. We always have a real negative eigenvalue $\lambda_1$. As shown in Appendix~\ref{A2}, for both $g_\bathw$- and $g_\bathc$-coupling systems, there are two possible distributions of the eigenvalues if there are not any LEPs: (a)~a zero eigenvalue and four distinct real eigenvalues, as exemplified in Fig.~\ref{nonEP1}; (b)~a zero eigenvalue, two distinct real eigenvalues and a pair of complex conjugate eigenvalues, as exemplified in Fig.~\ref{nonEP2}. We then have a second-order LEP when the pair of complex conjugate eigenvalues coalesces, as exemplified in Fig.~\ref{EP2}. We further have a third-order LEP when one real eigenvalue and the pair of complex conjugate eigenvalues coalesce, as exemplified in Fig.~\ref{EP3}. Note that there are various sequences of the eigenvalues in Fig.~\ref{Eigs}; for example, the real and negative eigenvalue $\lambda_1$ may be greater than the other eigenvalues.


\subsection{Conditions for the third-order LEP}
\label{Sec2-2}

Based on the derivation of the cubic polynomial equation in Appendix~\ref{A2}, we analyze the discriminant of the cubic polynomial equation, finding two relations among the parameters when $F_3(\lambda)$ has a triple root $\lambda^{(3)}$. From the two relations, we can reduce the two parameters $g_\alpha$ and $\gamma^+$, making them dependent on the other two parameters $\gamma^-$ and $\omega_\alpha$ as in Eq.~(\ref{relationEP3}). Therefore, the system only depends on the two parameters $\gamma^-$ and $\omega_\alpha$ when the triple root of  $F_3(\lambda)$ appears. We set the parameters $g_\alpha^{(3)}$ and ${\gamma^+}^{(3)}$ for the triple root $\lambda^{(3)}$ to appear as in Eq.~(\ref{relationEP3}) in the following analysis:
\begin{align}
\label{EP3Para1}
&g_\alpha^{(3)}(\omega_\alpha) =\sqrt{2} \omega_\alpha,\\
\label{EP3Para2}
&{\gamma^+}^{(3)}(\omega_\alpha, \gamma^-) = -\gamma^- + 6\sqrt{3} \omega_\alpha.
\end{align}

We can check that there is only one solution of the eigenvector $\rho_{5\times 5}^{(3)}$ for the triple root $\lambda^{(3)}$, and hence we confirm that the eigensystem is at a third-order exceptional point. For the $g_\bathw$-coupling system, the triple root is given by
\begin{align}
\lambda^{(3)}(\omega_\bathw, \gamma^-) = \gamma^- - 10\sqrt{3}\omega_\bathw,
\end{align}
while for the $g_\bathc$-coupling system, it is given by
\begin{align}
\lambda^{(3)}(\omega_\bathc, \gamma^-) = -\gamma^- - 4\sqrt{3}\omega_\bathc.
\end{align}

\subsection{Eigenvectors}
\label{Sec2-3}
At non-LEP points, the right eigenvector $\tilde{\rho}_j$ of each eigenvalue $\tilde{\lambda}_j$ can be normalized by its own left eigenvector $\tilde{\sigma}_j$, and it is bi-orthogonal to the left eigenvectors $\tilde{\sigma}_i$ of other eigenvalues $\tilde{\lambda}_i$:
\begin{align}
&\tilde{\sigma}_j (\hat{\mathcal{L}}_{5\times 5}-\tilde{\lambda}_j \mathbb{I}_{5})=0,\\ 
&(\hat{\mathcal{L}}_{5\times 5}-\tilde{\lambda}_j \mathbb{I}_{5}) \tilde{\rho}_j=0.\\
&\tilde{\sigma}_i \tilde{\rho}_j = \delta_{i,j}.
\end{align}

Here and hereafter, we let the tilder mark indicate that the quantity is given at a non-LEP.
The system state $\tilde{\rho}(t)$ away from the exceptional point can be thereby written as a weighted sum of all eigenvectors with exponentially decaying factors, except for the steady state $\tilde{\rho}_\text{ss}$ with the zero eigenvalue:
\begin{align}
&\tilde{\rho}(t) =\tilde{c}_{\textrm{ss}} \tilde{\rho}_{\textrm{ss}} + \sum_{j=1}^4 \tilde{c}_j e^{\tilde{\lambda}_j t}\tilde{\rho}_{j},\\
&\tilde{c}_j = \tilde{\sigma}_j \tilde{\rho}(0),
\end{align}
where we let the left and right eigenstates of the zero eigenvalue denoted by $\tilde{\sigma}_{\textrm{ss}}$ and $\tilde{\rho}_{\textrm{ss}}$, while those of the other eigenvalues $\tilde{\lambda}_j$ by $\tilde{\sigma}_j$ and $\tilde{\rho}_j$ with $1 \leq j \leq 4$. Note that the real parts of $\tilde{\lambda}_j$ are always negative, as is indicated in Fig.~\ref{Eigs}.

At the LEP, not just eigenvalues but eigenvectors collapse into one.
It is also evident that the orthogonality between the left and right eigenvectors is broken. In fact, the left eigenvector $\sigma_j$ and the right eigenvector $\rho_j$ of the eigenvalue $\lambda_j$ become orthogonal to each other at the exceptional points. By generalizing the eigensystem at the third-order LEP as described in Appendix \ref{A3}, we follow the Jordan chain to derive the three sets of pseudo-left and pseudo-right eigenvectors $\sigma_{EP}^{(n)}$ and $\rho_{EP}^{(n)}$ for $n=1,2,3$, out of which only $\sigma^{(3)}_{EP}$ and $\rho^{(1)}_{EP}$ are the true eigenvectors.
The system state $\rho(t)$ at the exceptional point can be written as a weighted sum of the right eigenvectors $\rho_j$ with exponentially decaying factors, with extra factors $t$ and $t^2$:
\begin{align}
\label{rhoEP3}
&\rho(t) = c_{\textrm{ss}} \rho_{\textrm{ss}} + c_1 e^{\lambda_1 t} \rho_{1} +\left [ \left(c_2 + c_3 t + c_4 \frac{t^2}{2}\right)\rho_{EP}^{(1)} + (c_3 + c_4 t )\rho_{EP}^{(2)} +  c_4 \rho_{EP}^{(3)} \right]e^{\lambda^{(3)}t},
\end{align}
where
\begin{align*}
&c_j = \sigma_j\rho(0) \text{ for } j=\textrm{ss},1;\quad c_i = \sigma_{EP}^{(i-1)} \rho(0) \text{ for } i=2,3,4.
\end{align*}

\section{Definitions of Heat}
\label{Sec3}
A quantum thermal machine can operate as three types of devices: (i)~engine $Q_\bathh>0$, $Q_\bathc<0$, $Q_\bathw<0$; (ii)~refrigerator $Q_\bathh<0$, $Q_\bathc>0$, $Q_\bathw>0$; and (iii)~heater $Q_\bathh>0$, $Q_\bathc<0$, $Q_\bathw>0$. We define the direction of energy flow as one from the external bath to the internal system; hence $Q_\alpha>0$ means that the internal system absorbs heat from the bath $\alpha$ while $Q_\alpha<0$ means that the system releases heat to the bath $\alpha$ for $\alpha=\bathh, \bathw, \bathc$.
Note that the one-coupling three-level systems here cannot produce work, which means that they cannot operate as a real engine, though the energy flows are similar to the engine in case (i). We hereafter focus on the case~(ii) of the refrigerator. We define the coefficient of performance (COP) for the refrigerator as $Q_\bathc/Q_\bathw$.

\subsection{instantaneous Heat Current at Time $t$}
The energy of the system is given by $E(t)=\tr [\hat{H}_\textrm{s} \rho(t)]$. We can thereby define the instantaneous energy change at time $t$ from the bath to the internal system using the Liouvillian superoperator:
\begin{align}
	\dot{E}(t) = \tr[\hat{H}_\textrm{s} \dot{\rho}(t)] = \tr[\hat{H}_\textrm{s} \hat{\mathcal{L}}\rho(t)],
\end{align}
and hence we can define the heat current $\dot{Q}_\alpha$ using the dissipation terms as follows:
\begin{align}
	\dot{Q}_\alpha(t) = \tr[\hat{H}_\textrm{s} \mathcal{L}_\alpha \rho(t)],
\end{align}
where
\begin{align}
	\hat{\mathcal{L}}_\alpha &= \gamma^+\hat{\mathcal{D}}[\ketg \bra{\alpha}] + \gamma^- \hat{\mathcal{D}}[\ket{\alpha} \brag] \text{ for } \alpha=\bathh, \bathc,\\
	\hat{\mathcal{L}}_\bathw&= \gamma^+\hat{\mathcal{D}}[\ketc \brah] + \gamma^- \hat{\mathcal{D}}[\keth \brac].
\end{align}
Using the Liouvillian superoperator in Eq.~(\ref{L9}), we can explicitly write down the non-equilibrium heat current $\dot{Q}_\alpha (t)$ at time $t$ in the following forms:
\begin{align}
	\label{Qct}
	\dot{Q}_c(t) = \omega_\bathc [\gamma^- \rho_\textrm{gg}(t) -\gamma^+ \rho_\textrm{cc}(t))-\frac{1}{2} \big(g_\bathw \gamma^+(\rho_\textrm{ch}(t)+\rho_\textrm{hc}(t))+g_\bathc(\gamma^++\gamma^-](\rho_\textrm{gc}(t)+\rho_\textrm{cg}(t))\big),\\
	\label{Qwt}
	\dot{Q}_w(t) = \omega_\bathw [\gamma^- \rho_\textrm{cc}(t) -\gamma^+ \rho_\textrm{hh}(t))-\frac{1}{2} \big(g_\bathc \gamma^-(\rho_\textrm{gc}(t)+\rho_\textrm{cg}(t))+g_\bathw(\gamma^++\gamma^-](\rho_\textrm{ch}(t)+\rho_\textrm{hc}(t))\big),\\
	\label{Qht}
	\dot{Q}_h(t) = (\omega_\bathc+\omega_\bathw) [\gamma^- \rho_\textrm{gg}(t) -\gamma^+ \rho_\textrm{hh}(t))-\frac{1}{2} \big(g_\bathw \gamma^+(\rho_\textrm{ch}(t)+\rho_\textrm{hc}(t)]+g_\bathc \gamma^-(\rho_\textrm{gc}(t)+\rho_\textrm{cg}(t))\big).
\end{align}
Hence the instantaneous non-equilibrium COP $\eta_\textrm{inst}(t)$ at time $t$ is define as the ratio of heat current $\dot{Q}_\bathc(t)$ of the cold bath to $\dot{Q}_\bathw(t)$ of the work bath:
\begin{align}
	\label{COPint}
	\eta_\textrm{inst}(t) = \frac{\dot{Q}_\bathc(t)}{\dot{Q}_\bathw(t)}.
\end{align} 

\subsection{Entire Heat Absorption in Time $t$}
The accumulated heat change $Q_\alpha(t)$ from the initial state $\rho(0)$ to the state $\rho(t)$ with system evolution time $t$ is given by
\begin{align}
	\label{QB}
	Q_\alpha(t) =  \int_{0}^{t} \dot{Q}_\alpha(\tau) \,d\tau.
\end{align}
We can also define the COP $\eta_\textrm{accum}(t)$ as the ratio of accumulated heat extraction $Q_\bathc(t)$ of the cold bath to $Q_\bathw(t)$ of the work bath:
\begin{align}
	\label{COPtot}
	\eta_\textrm{accum}(t) = \frac{Q_\bathc(t)}{Q_\bathw(t)}.
\end{align} 



\section{Critical Damping}
\label{Sec4}
In the present section, by comparing the damping speed between the third-order LEP states and the non-LEP states, we demonstrate the critical damping of the system states and heat current at the third-order LEP. 

We compare the dynamics at the LEP and away from the LEP as follows. For the former, we specify $\gamma^-$ and $\omega_\alpha$ with $\alpha=\bathc$ or $\bathw$, set $g_\alpha$ and $\gamma^+$ to the values given by $\gamma^-$ and $\omega_\alpha$ as in Eqs.~(\ref{EP3Para1}) and (\ref{EP3Para2}), and examine the equiribration dynamics to the stationary states and heat currents. For the latter, we specify the parameter values $\gamma^+$, $\gamma^-$, $\omega_\alpha$, $g_\alpha$ away from the LEP and examine the dynamics to the corresponding stationary state and heat current $\dot{Q}_\bathc(t)$ from the cold bath to the internal system.

\subsection{Critical Damping of System State}
\label{Sec4-1}
To quantitatively compare the critical damping at the third-order LEP to the damping of the near-LEP states, we introduce a quantity $\mathcal{R}_\textrm{s}$, which is the ratio between the trace distance of the LEP state and that of the non-EP state:
\begin{align}
\label{Rst}
\mathcal{R}_\textrm{s}(t) =& \frac{||\rho(t)-\rho_{ss}||_1}{||\tilde{\rho}(t)-\tilde{\rho}_{ss}||_1}\\
=&\frac{|| c_1 e^{\lambda_1t} \rho_1 + \bigl[ (c_2+c_3 t + c_4 \frac{t^2}{2}) \rho_{EP}^{(1)} +(c_3+c_4 t ) \rho_{EP}^{(2)} +c_4\rho_{EP}^{(3)} \bigr]e^{\lambda^{(3)}t}
||_1}{|| \tilde{c}_1 e^{\tilde{\lambda}_1t} \tilde{\rho}_1 + \tilde{c}_2 e^{\tilde{\lambda}_2t} \tilde{\rho}_2+\tilde{c}_3 e^{\tilde{\lambda}_3t}\tilde{\rho}_3 + \tilde{c}_4 e^{\tilde{\lambda}_4t}\tilde{\rho}_4 ||_1},
\end{align}
where $||\cdot||_1$ is the 1-norm. The critical damping is characterized by the following two aspects. (A)~After s short duration $\tau_\textrm{s}$, the trace distance at the third-order LEP becomes shorter than the one for non-LEP, and hence the ratio $\mathcal{R}_\textrm{s}(t)$ between LEP-state $\rho(t)$ and away-LEP state $\tilde{\rho}(t)$ is smaller than unity. (B)~After a long duration $t\rightarrow \infty$, the ratio $\mathcal{R}_\textrm{s}(t)$ approaches zero:
\begin{align}
\label{RsCD}
\mathcal{R}_\textrm{s}(t)<1\text{ for }{t>\tau_\textrm{s}} \text{ and }\lim_{t\rightarrow  \infty}\mathcal{R}_\textrm{s}(t)\rightarrow 0.
\end{align}
\begin{figure}
	\centering
	\begin{subfigure}{0.45\textwidth}
		\includegraphics[width=1.\textwidth]{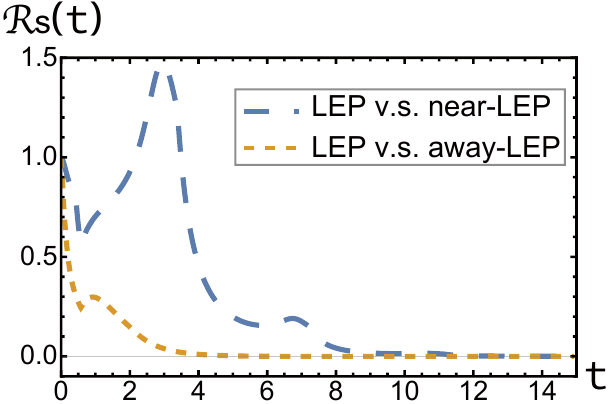}
		\caption{$g_\bathw$-coupling}
		\label{gwRs}
	\end{subfigure}
	\begin{subfigure}{0.45\textwidth}
		\includegraphics[width=1.0\textwidth]{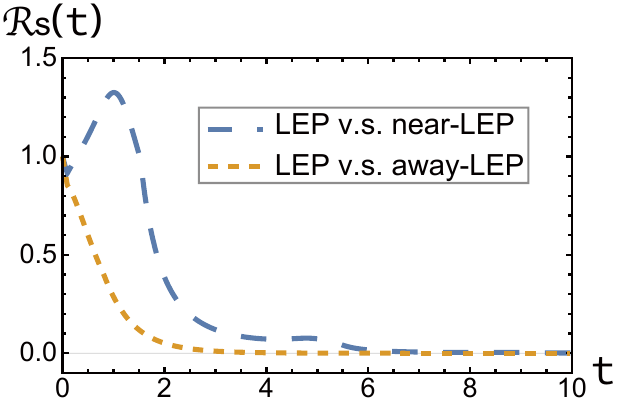}
		\caption{$g_\bathc$-coupling}
		\label{gcRs}
	\end{subfigure}
	\caption{Time-dependence of the ratio $\mathcal{R}_\textrm{s}(t)$ for (a)~the $g_\bathw$-coupling system and (b)~the $g_\bathc$-coupling system. We specify the energy levels as $\omega_\bathw=\omega_\bathc=1$ as the energy unit and set the dissipation rate $\gamma^+$ as in Eq.~(\ref{EP3Para2}) for all states. Following the condition~(\ref{EP3Para1}), we set the coupling strength $g_\alpha=(\sqrt{2}/2)\omega_\alpha$ at the third-order LEP. We choose the values of the coupling strength as $g_\alpha=1.314$ for the state near LEP (``near-LEP") and $g_\alpha=0.414$ for the state away from LEP (``away-LEP"). The parameters $\{\gamma^-,c_1,c_2,c_3,c_4\}$ are $\{7.159, 0, 0.125, 0.1,0\}$ for (a)~the $g_\bathw$-coupling system while $\{5.857, 0.127,0.1,0, 0\}$ for (b)~the $g_\bathc$-coupling system}
	\label{Rs}
\end{figure}

In Fig.~\ref{Rs}, we show the dynamics of $\mathcal{R}_\textrm{s}(t)$. 
For the numerator of $\mathcal{R}_\textrm{s}(t)$, we set $\gamma^+$ and $g_\alpha$ as in Eqs.~(\ref{EP3Para1}) and (\ref{EP3Para2}) depending on $\gamma^-$ and $\omega_\alpha$ to make the state at the third-order LEP. For the denominator, we specified the values of $\gamma^+$, $\gamma^-$, and $\omega_\alpha$ the same as in the numerator, and adjusted $g_\alpha$ to change the distance of the non-LEP states from the third-order LEP. When $g_\alpha$ approaches the one in (\ref{EP3Para2}), the state in the denominator gets closer to the third-order LEP state. 
In Fig.~\ref{Rs}, we chose $g_\alpha=(\sqrt{2}-0.1)\omega_\alpha\approx1.314$ to make the state near the third-order LEP (``near-LEP") and $g_\alpha=(\sqrt{2}-1)\omega_\alpha\approx0.414$ to make the state away from the third-order LEP (``away-LEP"). Compared to both of the non-LEP states, the critical damping at the third-order LEP is confirmed. Compared to the ``away-LEP" state, the ratio $\mathcal{R}_\textrm{s}(t)$ of ``near-LEP" state damps strongly. In other words, when the system state gets closer to the LEP, the damping speed becomes rapid.

For the choice of the parameters $\gamma^-$ and $\omega_\alpha$, we note the following fact. As detailed in Appendix~\ref{A4} under the condition $\lambda_1<\lambda^{(3)}<0$, the LEP dominates the non-equilibrium dynamics in Eq.~(\ref{Rst}), and hence the system always achieves the critical damping. We can rewrite the condition $\lambda_1<\lambda^{(3)}<0$ as a relation for the $g_\bathw$- and $g_\bathc$-coupling systems, repectively:
\begin{align}
\label{gwCDcondition}
\text{$g_\bathw$-coupling: } 2\sqrt{3}\omega_\bathw<\gamma^-<6\sqrt{3}\omega_\bathw,\\
\label{gcCDcondition}
\text{$g_\bathc$-coupling: } \gamma^-<4\sqrt{3}\omega_\bathc.
\end{align}
Here and hereafter, we fix the energy unit of the dissipation rate $\gamma^-$ to the energy level $\omega_\alpha$ ($\alpha=\bathh, \bathc$). 
In Fig.~\ref{Rs}, we fixed the dissipation rate $\gamma^-$ and the coefficients $c_{1,2,3,4}$ of the initial state so that Eqs.~(\ref{gwCDcondition}) and (\ref{gcCDcondition}) may be satisfied. As detailed in Sec.~\ref{Sec5-1} and Appendix~\ref{A5}, the target values of the parameters $\{\gamma^-,c_1,c_2,c_3,c_4\}$ should allow the systems to achieve a better performance and the optimal initial non-equilibrium heat current $\dot{Q}_\bathc^^i=\dot{Q}_\bathc(0)$ from the cold bath; we thereby chose specific values $\{7.159, 0, 0.125, 0.1,0\}$ for $g_\bathw$-coupling system and $\{5.857, 0.127,0.1,0, 0\}$ for $g_\bathc$-coupling system.

\subsection{Critical Damping of Heat Current}
\label{Sec4-2}
Since all of the heat current $\dot{Q}_\alpha(t)$ in Eqs.~(\ref{Qct})--(\ref{Qht}) from each bath $\alpha=\bathh,\bathc,\bathw$ are given by the weighted sum of the system state $\rho(t)$, in the present section, we analyze $\dot{Q}_\bathc(t)$ as an example to demonstrate the critical damping of the heat currents. Other heat currents should have a similar damping as $\dot{Q}_\bathc(t)$.

Similarly to the system state, for which we used the trace norm to measure the distance between the steady state and the non-equilibrium states, we use the absolute value of the difference between the heat current at the non-equilibrium state and the equilibrium heat current here to quantitatively compare the critical damping of the heat current:
\begin{align}
\label{Rct}
\mathcal{R}_\bathc(t) =\frac{\left |\dot{Q}_\bathc(t)-\dot{Q}_\bathc^\textrm{ss}\right |}{\left |\dot{\tilde{Q}}_\bathc(t)-\dot{\tilde{Q}}_\bathc^\textrm{ss}\right |}.
\end{align}
The heat currents $\dot{Q}_\bathc(t)$ and $\dot{Q}_\bathc^\textrm{ss}$ are given by the parameters making the system at the third-order LEP, while the heat currents $\dot{\tilde{Q}}_\bathc(t)$ and $\dot{\tilde{Q}}_\bathc^\textrm{ss}$  are given by the parameters making the system at non-LEP states.

We can observe the critical damping in the following two aspects: (A)~After s short duration $\tau_\bathc$, the diffference of the heat current at the third-order LEP becomes smaller than the one for non-LEP, and hence the ratio $\mathcal{R}_\bathc(t)$ is smaller than unity. (B)~After a long duration $t\rightarrow \infty$, the ratio $\mathcal{R}_\bathc(t)$ approaches zero:
\begin{align}
\label{RcCD}
\mathcal{R}_\bathc(t)<1\text{ for }{t>\tau_\bathc} \text{ and }\lim_{t\rightarrow  \infty}\mathcal{R}_\bathc(t)\rightarrow 0.
\end{align}
\begin{figure}
	\centering
	\begin{subfigure}{0.4\textwidth}
		\includegraphics[width=1.\textwidth]{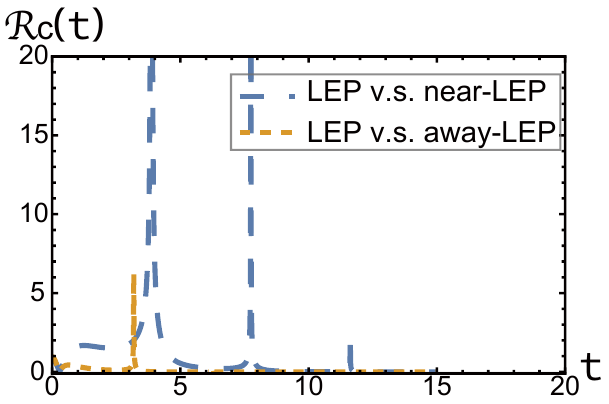}
		\caption{$g_\bathw$-coupling}
		\label{gwRc}
	\end{subfigure}
	\begin{subfigure}{0.4\textwidth}
		\includegraphics[width=1.\textwidth]{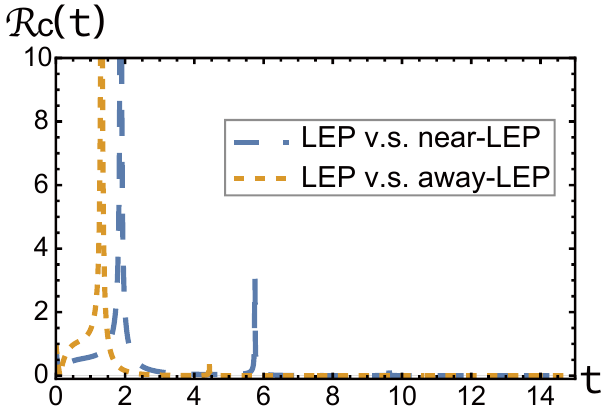}
		\caption{$g_\bathc$-coupling}
		\label{gcRc}
	\end{subfigure}
	\caption{Time-dependence of the ratio $\mathcal{R}_\bathc(t)$ for (a)~the $g_\bathw$-coupling system and (b)~the $g_\bathc$-coupling system. The other parameters $\{\gamma^-,c_1,c_2,c_3,c_4\}$ are $\{7.159, 0, 0.125, 0.1,0\}$ for (a)~the $g_\bathw$-coupling system while $\{5.857, 0.127,0.1,0, 0\}$ for (b)~the $g_\bathc$-coupling system. }
	\label{Rc}
\end{figure}

In Fig.~\ref{Rc}, for the analysis of the ratio $\mathcal{R}_\bathc(t)$ for the heat current, we chose the same values of parameters as the analysis of the system state in Fig.~\ref{Rs}. 
We observed the critical damping of the heat current $\dot{Q}_\bathc(t)$ with the parameters that achieve a better performance and the optimal initial non-equilibrium heat current from the cold bath $\dot{Q}_\bathc(0)$, as detailed in Sec.~\ref{Sec5-1} and Appendix~\ref{A5}. The oscillation in a short time in Fig.~\ref{Rc} is caused by the oscillation of the denominator.  Compared to the ``away-LEP" state, the oscillation of the ``near-LEP" state is stronger and persists longer. Similarly to the damping of the system state, the damping of the heat current also becomes faster when the system gets closer to the third-order LEP. 
Since the heat currents are the weighted sum of the system states, the critical damping of the heat currents also confirms the critical damping of the system states.

\section{Performance at the Third-Order LEPs}
\label{Sec5}
After the system state converges to the steady state, the dynamics becomes the same as the conventional research~\cite{PhysRevE.97.052145, PhysRevE.100.062112, PhysRevE.101.062121, PhysRevE.92.012136, Correa2014} on the absorption refrigerator at the equilibrium state. However, the thermal dynamics of the system can be different in the non-equilibrium process before the system converges to the steady state. The entire heat exchange between the internal system and the external bath in the non-equilibrium process may be non-negligible.

In the present section, we analyze the influence of the initial state and search for proper initial states for the $g_\bathw$-coupling- and $g_\bathc$-coupling systems to achieve better performance. We confirm better performance in both systems by comparing heat current $\dot{Q}_\alpha(t)$, entire heat extraction $Q_\alpha(t)$, the instantaneous COP $\eta_\textrm{inst}$ and the accumulated $\eta_\textrm{accum}$ between the third-order LEP state and the steady state.

\subsection{Initial State}
\label{Sec5-1}
The heat exchange starts from the initial heat current $\dot{Q}_\alpha(0)$ given by the initial state $\rho(0)$ and converges to the equilibrium heat current $\dot{Q}_\alpha^\textrm{ss}$. The choice of the initial state is crucial for the non-equilibrium dynamics before equilibrium. In the present section, we reveal the influence of the initial state and search for a proper initial state for our purposes.

A refrigerator transfers heat from the cold bath $\bathc$ to the hot bath $\bathh$, supported by the work bath $\bathw$ in our model. The COP is the ratio of the heat absorption from the cold bath to the work extraction from the work component, as in Eqs.~(\ref{COPint}) and (\ref{COPtot}). We require two conditions for the absorption refrigerator to achieve better performance: 
(I) The internal system absorb more heat from the cold bath and releases more heat to the hot bath than at the steady state $\dot{Q}_\bathh(t)<\dot{Q}_\bathh^\textrm{ss}<0$ and $0<\dot{Q}_\bathc^\textrm{ss}<\dot{Q}_\bathc(t)$; 
(II) The COP is higher as $0<\eta^\textrm{ss}<\eta_\textrm{inst}(t)$, which means the internal system absorbs more heat from the cold bath with less energy extraction from the work bath than in the steady state. 

At the third-order LEP, as shown in Sec.~\ref{Sec2} and Appendix~\ref{A1}, we generalize the eigensystem of the $5\times 5$ Liouvillian block $\mathcal{L}_{5\times 5}$ by left eigenvectors $\sigma_j$ and $\sigma_\textrm{EP}^{(i-1)}$ and right eigenvectors $\rho_j$ and $\rho_\textrm{EP}^{(i-1)}$ for $j=\textrm{ss}, 1$ and $i=2,3,4$. Any system state at the third-order LEP is given by the weighted sum of the right eigenvectors as in Eq.~(\ref{rhoEP3}), and hence the initial state $\rho(0)$ is also given by
\begin{align}
\label{rho0}
&\rho(0) = \rho_{\textrm{ss}} + c_1 \rho_{1} +c_2 \rho_{EP}^{(1)} + c_3 \rho_{EP}^{(2)} +  c_4 \rho_{EP}^{(3)}.
\end{align}

We specify the energy levels $\omega_\alpha$ ($\alpha=\bathc, \bathw$) of the system as the energy unit, and thereby fix the coupling strength $g_\alpha$ and one of the dissipation rates $\gamma^+$ as in Eqs.~(\ref{EP3Para1}) and (\ref{EP3Para2}). The system is hence controlled only by the dissipation rate $\gamma^-$ and the coefficients $c_k$ $(k=1,2,3,4)$ of the initial state. 

As detailed in Appendix~\ref{A5}, fixing the systems to the case of the critical damping as in Eqs.~(\ref{gwCDcondition})--(\ref{gcCDcondition}), we focus on the initial heat current $\dot{Q}_\alpha^i=\dot{Q}_\alpha(0)$ that achieves a better performance. 
First, we analyze the influence of each coefficient $c_k$ on the initial non-equilibrium heat current $\dot{Q}_\alpha^i$ and search for the proper initial states that improve the performance of the refrigerator. The coefficients only influence the non-equilibrium heat current $\dot{Q}_\alpha^i$ without changing the equilibrium heat current $\dot{Q}_\alpha^\textrm{ss}$. Based on the role of each coefficient, we demonstrate the influence of the non-equilibrium term on the non-equilibrium heat currents and analytically search for better performance as constraints (I) and (II) by adjusting the coefficients.

In addition, we adjust the dissipation parameters $\gamma^-$ and make the initial non-equilibrium heat currents $\dot{Q}_\alpha^i$ not only satisfy the two constraints of a better performance but also obtain the optimal $\dot{Q}_\bathc^i$. The dissipation $\gamma^-$ influences not only the initial non-equilibrium heat current $\dot{Q}_\alpha^i$ but also the equilibrium heat current $\dot{Q}_\alpha^\textrm{ss}$. Focusing on the optimal non-equilibrium initial heat current $\dot{Q}_\bathc^i$, after fixing the coefficients $c_k$ of the initial state, we search for a value of the dissipation rate $\gamma^-$ that achieves it.

As detailed in Appendix~\ref{A5}, the initial non-equilibrium heat current $\dot{Q}_\bathc^i$ always increases as we increase $c_3$ in the $g_\bathw$-coupling system while $c_1$ and $c_2$ in the $g_\bathc$-coupling system. After fixing them, we next search for the optimal value of the initial non-equilibrium heat current $\dot{Q}_\bathc^i$. 
For the $g_\bathw$-coupling system, if we set $c_3=0.1$, we find the highest heat current $\dot{Q}_\bathc^i=1.330$ numerically for $\{\gamma^-,c_1,c_2,c_4\}=\{7.159, 0, 0.125, 0\}$. For the $g_\bathc$-coupling system, if we set $c_2=0.1$, we numerically find that the optimal heat current $1.93$ is given by $\{\gamma^-,c_1,c_3,c_4\}=\{5.857, 0.127,0, 0\}$. We chose these sets of parameter values in all the numerical calculations in the present paper.

\subsection{Heat and COP}
\label{Sec5-2}

Starting from the initial states given in the previous section, in the present section, we compare the heat current $\dot{Q}_\alpha(t)$, the entire heat extraction $Q_\alpha(t)$, the instantaneous COP $\eta_\textrm{inst}(t)$ and the accumulated COP $\eta_\textrm{accum}(t)$ between the third-order LEP system and the equilibrium system.

\begin{figure*}
	\centering
	\begin{subfigure}{0.4\textwidth}
		\includegraphics[width=1\textwidth]{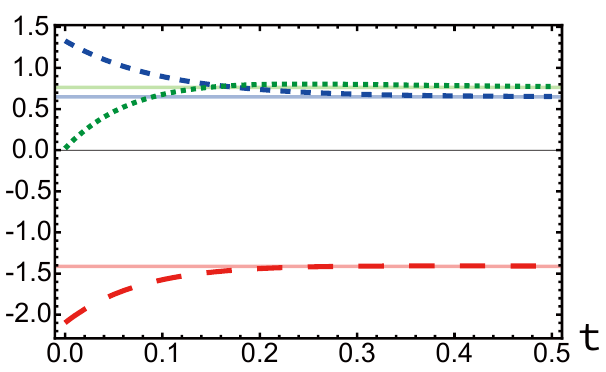}
		\caption{$g_\bathw$-coupling}	
		\label{gwdotQ}
	\end{subfigure}
	\begin{subfigure}{0.4\textwidth}
		\includegraphics[width=1\textwidth]{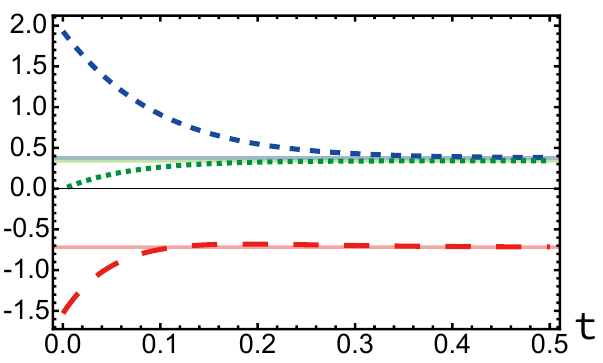}
		\caption{$g_\bathc$-coupling}
		\label{gcdotQ}
	\end{subfigure}
	\begin{subfigure}{0.15\textwidth}
		\centering
		\includegraphics[width=0.7\textwidth]{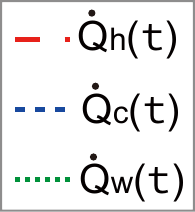}
		\caption*{}
	\end{subfigure}
	\caption{Time-dependence of the $\dot{Q}_\alpha(t)$ $(\alpha=\bathc, \bathw, \bathh)$ in (a)~$g_\bathw$-coupling system with and (b)~$g_\bathc$-coupling system . As conditions~(\ref{EP3Para1}) and (\ref{EP3Para2}), we set $\gamma^+ = 6\sqrt{3}\omega_\alpha-\gamma^-$, and $g_\alpha=\sqrt{2}\omega_\alpha$ and specify the energy levels to $\omega_\bathw=\omega_\bathc=1$ as the energy unit. The horizontal lines in the figures are the equilibrium heat currents $\dot{Q}_\alpha^\textrm{ss}$. As in Sec.~\ref{Sec5-1} and Appendix~\ref{A5}, we set the parameters $\{\gamma^-, c_1, c_2, c_3, c_4\}$ of (a)~$g_\bathw$-coupling system to $\{7.159, 0, 0.125, 0.1, 0\}$ and of (b)~$g_\bathc$-coupling system to $\{5.857, 0.127,0.1,0, 0\}$.}
	\label{dotQ}
\end{figure*}
As shown in Fig.~\ref{dotQ}, the heat exchanges in both the $g_\bathw$-coupling and $g_\bathc$-coupling systems start from the initial value given by the coefficients that we set and critically damp to the equilibrium heat currents, which confirms the critical damping of heat current in Sec.~\ref{Sec4-2}.
Both systems satisfy the requirements for better performance, (I)~transferring more heat from the cold bath to the hot bath and (II)~absorbing more heat from the cold bath with less energy extraction from the work bath than at the steady state. Figure~\ref {dotQ} confirms that a better performance is achieved in the non-equilibrium process compared to the equilibrium system. 


\begin{figure*}
	\centering
	\begin{subfigure}{0.4\textwidth}
		\includegraphics[width=1\textwidth]{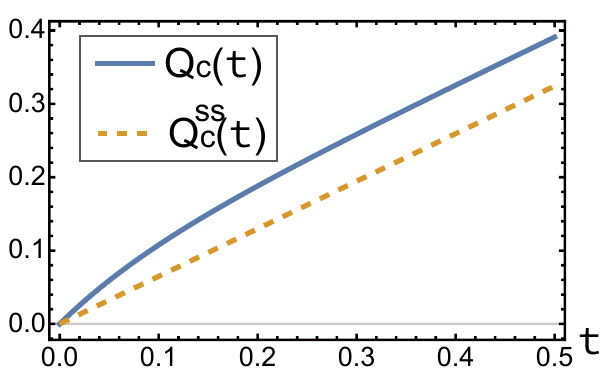}
		\caption{$g_\bathw$-coupling}	
		\label{gwQc}
	\end{subfigure}
	\begin{subfigure}{0.4\textwidth}
		\includegraphics[width=1\textwidth]{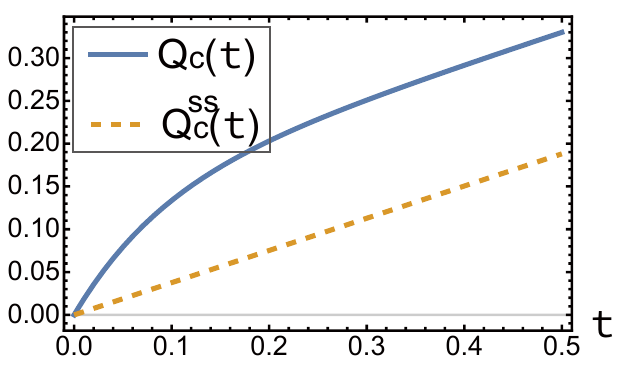}
		\caption{$g_\bathc$-coupling}
		\label{gcQc}
	\end{subfigure}
	\caption{Time-dependence of the $Q_\alpha(t)$ and $Q_\alpha^\textrm{ss}(t)$ in (a)~$g_\bathw$-coupling system and (b)~$g_\bathc$-coupling system. We set the parameters the same as in Fig.~\ref{dotQ}. As in Sec.~\ref{Sec5-1} and Appendix~\ref{A5}, we set the parameters $\{\gamma^-, c_1, c_2, c_3, c_4\}$ of a)~$g_\bathw$-coupling system to $\{7.159, 0, 0.125, 0.1, 0\}$ and of (b)~$g_\bathc$-coupling system to $\{5.857, 0.127,0.1,0, 0\}$.}
	\label{Qc}
\end{figure*}

Figure~\ref{Qc} shows that the entire heat absorption $Q_\bathc(t)$ converges to a value different from the equilibrium one $Q_\bathc^\textrm{ss}$, compared to the instantaneous heat currents $\dot{Q}_\alpha(t)$ that converges to the equilibrium ones $\dot{Q}_\alpha^\textrm{ss}$ after the non-equilibrium process.

\begin{figure*}
	\centering
	\begin{subfigure}{0.4\textwidth}
		\includegraphics[width=1\textwidth]{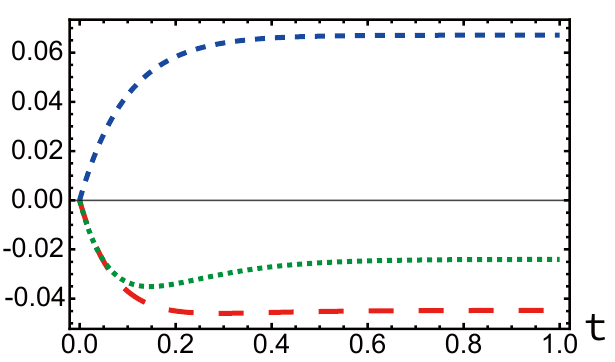}
		\caption{$g_\bathw$-coupling}	
		\label{gwQ}
	\end{subfigure}
	\begin{subfigure}{0.4\textwidth}
		\includegraphics[width=1\textwidth]{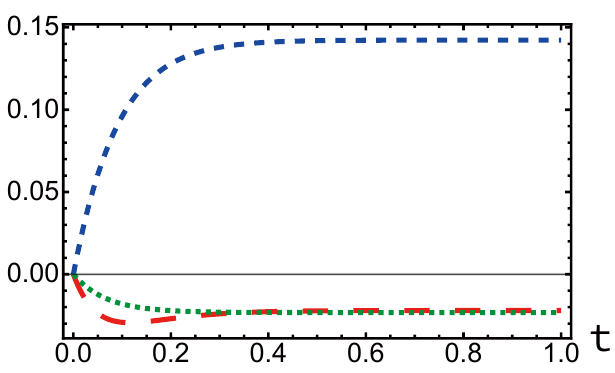}
		\caption{$g_\bathc$-coupling}
		\label{gcQ}
	\end{subfigure}
	\begin{subfigure}{0.15\textwidth}
		\centering
		\includegraphics[width=1\textwidth]{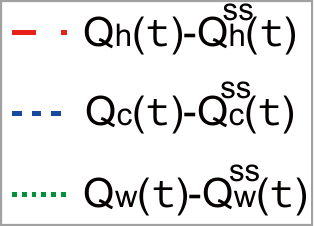}
		\caption*{}
	\end{subfigure}
	\caption{Time-dependence of the $Q_\alpha(t)-Q_\alpha^\textrm{ss}(t)$ $(\alpha=\bathc, \bathw, \bathh)$ in (a)~$g_\bathw$-coupling system and (b)~$g_\bathc$-coupling system. We set the parameters the same as in Fig.~\ref{dotQ}. As in Sec.~\ref{Sec5-1} and Appendix~\ref{A5}, we set the parameters $\{\gamma^-, c_1, c_2, c_3, c_4\}$ of (a)~$g_\bathw$-coupling system to $\{7.159, 0, 0.125, 0.1, 0\}$ and of (b)~$g_\bathc$-coupling system to $\{5.857, 0.127,0.1,0, 0\}$.}
	\label{Q}
\end{figure*}

For comparison of the entire heat exchanges between the non-equilibrium process and the equilibrium process, we plot the energy differences $Q_\alpha(t)-Q_\alpha^\textrm{ss}(t)$ in Fig.~\ref{Q}, in which the nonlinear changes represent the non-equilibrium process. The energy differences approach specific values, which show the extra heat absorption from bath $\alpha$ in the entire non-equilibrium process compared to the equilibrium process. As shown in Fig.~\ref{Q}, the non-equilibrium process makes both systems transfer more heat from the cold bath to the hot bath with less energy cost from the work bath.

\begin{figure*}
	\centering
	\begin{subfigure}{0.4\textwidth}
		\includegraphics[width=1\textwidth]{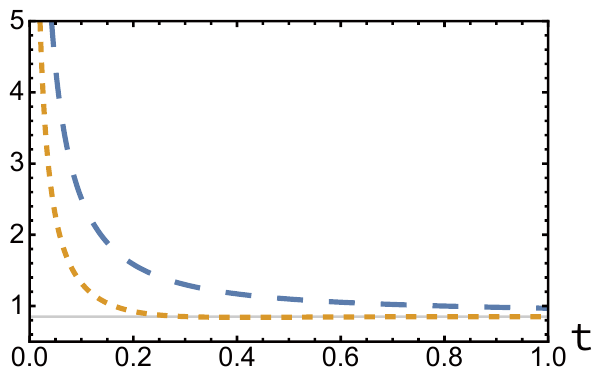}
		\caption{$g_\bathw$-coupling}
		\label{gwCOP}
	\end{subfigure}
	\begin{subfigure}{0.4\textwidth}
		\includegraphics[width=1\textwidth]{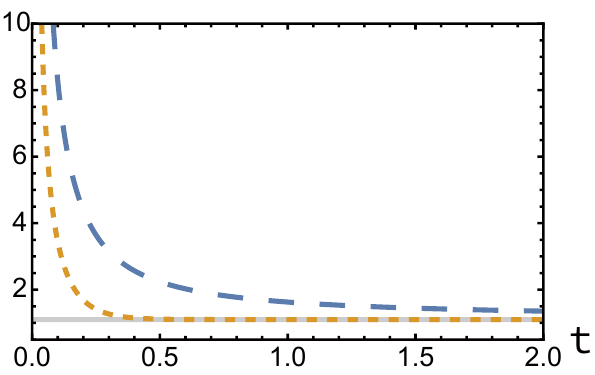}
		\caption{$g_\bathc$-coupling}
		\label{gcCOP}
	\end{subfigure}
	\begin{subfigure}{0.15\textwidth}
		\includegraphics[width=1\textwidth]{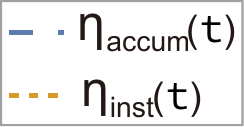}
		\caption*{ }
	\end{subfigure}
	\caption{Time-dependence of the COP $\eta_\textrm{inst}(t)$ of instantaneous heat current $\dot{Q}_\alpha(t)$ and the COP $\eta_\textrm{accum}(t)$ of entire heat absorption$Q_\alpha(t)$ in (a)~$g_\bathw$-coupling system and (b)~$g_\bathc$-coupling system. As in Sec.~\ref{Sec5-1} and Appendix~\ref{A5}, we set the parameters $\{\gamma^-, c_1, c_2, c_3, c_4\}$ of (a)~$g_\bathw$-coupling system to $\{7.159, 0, 0.125, 0.1, 0\}$ and of (b)~$g_\bathc$-coupling system to $\{5.857, 0.127,0.1,0, 0\}$.}
	\label{COP}
\end{figure*}
Figure~\ref{COP} shows that the instantaneous COP $\eta_\textrm{inst}(t)$ converges to the equilibrium COP $\eta_\textrm{inst}^\textrm{ss}$ after critical damping since the heat currents $\dot{Q}_\bathw(t)$ and $\dot{Q}_\bathc(t)$ converge to equilibrium. On the other hand, for the entire heat absorption $Q_\alpha(t)$, the COP $\eta_\textrm{accum}(t)$ approaches the equilibrium one $\eta_\textrm{accum}^\textrm{ss}$  later than $\eta_\textrm{inst}(t)$ appraoches $\eta_\textrm{inst}^\textrm{ss}$ since there are extra heat absorptions acummulated in the non-equilibrium process as shown in Fig.~\ref{Q}. 


\section{Conclusions and prospects}
\label{Conclusion}

In this work, we analyzed the dynamics of a quantum absorption refrigerator comprised of a qutrit internal system and three external heat baths. Focusing on the Liouvillian superoperator, we describe the quantum thermal machines in the context of non-Hermitian physics. Although the system Hamiltonian itself is still Hermitian, due to the influence of the interaction between the internal system and the external bath, the Liouvillian superoperator is non-Hermitian, which leads to a non-equilibrium process in the short-term evolution.

First, we analyzed the eigenvalue distributions of the Liouvillian superoperator and demonstrated the existence of the second- and third-order LEPs in the systems of two types of internal couplings. The dynamics of the systems with different types of couplings are similar. Focusing on the third-order LEP, we analyzed the dynamics during the non-equilibrium process from two points of view, namely the speed of damping from the initial state to the steady state and the influence of the non-equilibrium process on the performance of the quantum absorption refrigerator. 

For the damping speed, we derived the condition of the critical damping and demonstrated the critical damping of not only the system state but also the heat current. The non-equilibrium process is inevitable in an open quantum system, but it can be shortened by using the critical damping at LEPs if one wants models to achieve a steady state quickly and smoothly.

For the performance of the quantum absorption refrigerator in the non-equilibrium process, we analyzed the influence of various parameters on the non-equilibrium heat current and achieved more heat transfer from the cold bath to the hot bath with less energy cost from the work bath at the third-order LEP than at the steady state. We discovered that the system can achieve better performance in the non-equilibrium process at the third-order LEP than in the steady-state one.

The critical damping and the better performance happen under rather restricted conditions. As shown in Appendix~\ref{A4}, the relations~(\ref{gwCDcondition})--(\ref{gcCDcondition}) are required for our model to exhibit critical damping. In Sec.~\ref{Sec5-1} and Appendix~\ref{A5}, we explored the influence of the initial state and obtained the proper initial state for achieving better performance in our models.

All the results show the significance of considering the non-equilibrium process. The structure and the time evolution of most quantum thermal machines are based on open quantum systems, and hence the non-equilibrium process is inevitable. Searching for better or even the best conditions in the non-equilibrium process is important for future research on quantum thermal machines. 

\section{Acknowledgement}
J.G. acknowledges financial support from the WINGS-QSTEP program of the University of Tokyo. N.H.'s work was supported by JSPS KAKENHI Grant Numbers JP24K00545 and JP23K22411.

\appendix
\section{One-coupling Quantum Absorption Refrigerators}
\label{A1}

In order to reduce the complexity of the model in Sec.~\ref{Sec1}, in the present paper, we consider two types of one-coupling systems by restricting the two couplings $g_\bathw$ and $g_\bathc$ between neighboring energy levels to one. In both cases, the Liouvillian superoperator in Eq.~(\ref{L9}) can be rewritten to a block-diagonalized form with two $2 \times 2$ blocks and a $5\times 5$ block:
\begin{align}
\label{Lblock}
&\mathcal{L}_{9 \times 9} = \begin{bmatrix}
\mathcal{L}_{2 \times 2} &0 & 0\\
0 & \mathcal{L}_{2 \times 2}^* &0\\
0 &0 &\mathcal{L}_{5 \times 5}
\end{bmatrix}.
\end{align}

We show below that relevant parameters are $\{ \omega_\alpha, g_\alpha, \gamma^-, \gamma^+\}$ for $\alpha=\bathw,\bathc$ as specified at the end of Sec.~\ref{Sec2-2}.

\subsection{$g_\bathw$-coupling system: $g_\bathc=0$ and $g_\bathw \neq 0$}
\label{A1-1}
If we neglect the coupling $g_\bathc$ and keep only $g_\bathw$, the three blocks in Eq.~(\ref{Lblock}) read
\begin{align}
&\mathcal{L}_{2 \times 2} = \begin{bmatrix}
\mathrm{i}\omega_\bathc-\Gamma_\bathc & \mathrm{i}g_\mathrm{\bathw}\\
\mathrm{i}g_\mathrm{\bathw} &\mathrm{i}\omega_\bathh-\Gamma_\bathh
\end{bmatrix},\\
&\mathcal{L}_{5\times 5}=\begin{bmatrix}
-\gamma_h^--\gamma_\bathc^- &\gamma_\bathc^+ &0 &0 &\gamma_h^+\\
\gamma_\bathc^- &-\gamma_\bathc^+-\gamma_\mathrm{\bathw}^- & \mathrm{i}g_\mathrm{\bathw} &-\mathrm{i}g_\mathrm{\bathw} & \gamma_\bathw^+ \\
0 &\mathrm{i}g_\mathrm{\bathw} &\mathrm{i}\omega_\mathrm{\bathw}-\Gamma_\mathrm{\bathw} &0 &-\mathrm{i}g_\mathrm{\bathw}\\
0 &-\mathrm{i}g_\mathrm{\bathw} &0 &-\mathrm{i}\omega_\mathrm{\bathw}-\Gamma_\mathrm{\bathw} &\mathrm{i}g_\mathrm{\bathw}\\
\gamma_h^- &\gamma_\mathrm{\bathw}^- &-\mathrm{i}g_\mathrm{\bathw} &\mathrm{i}g_\mathrm{\bathw} &-\gamma_\mathrm{\bathw}^+-\gamma_h^+
\end{bmatrix}.
\end{align}
The basis of $\mathcal{L}_{2 \times 2}$ is $\begin{bmatrix}
 \rho_\textrm{gc} &\rho_\textrm{gh}
  \end{bmatrix}^T$, 
 that of $\mathcal{L}_{2 \times 2}^*$ is $\begin{bmatrix} 
 \rho_\textrm{cg} &\rho_\textrm{hg}
 \end{bmatrix}^T$, and that of $B_{5\times 5}$ is $\begin{bmatrix} \rho_\textrm{gg} &\rho_\textrm{cc} &\rho_\textrm{ch} &\rho_\textrm{hc} &\rho_\textrm{hh} \end{bmatrix}^T$.
 If we further set $\gamma^+=\gamma_\bathc^+=\gamma_\bathh^+=\gamma_\bathw^+$ and $\gamma^-=\gamma_\bathc^-=\gamma_\bathh^-=\gamma_\bathw^-$, we have
 \begin{align}
 &\mathcal{L}_{2 \times 2} = \begin{bmatrix}
\mathrm{i}\omega_\bathc-\frac{1}{2}(3\gamma^- +\gamma^+) & \mathrm{i}g_\mathrm{\bathw}\\
\mathrm{i}g_\mathrm{\bathw} &\mathrm{i}\omega_\bathh-(\gamma^+ +\gamma^-)
\end{bmatrix},\\
 &\mathcal{L}_{5\times 5}=\begin{bmatrix}
-2\gamma^- &\gamma^+ &0 &0 &\gamma^+\\
\gamma^- &-\gamma^+ -\gamma^- & \mathrm{i}g_\mathrm{\bathw} &-\mathrm{i}g_\mathrm{\bathw} & \gamma^+ \\
0 &\mathrm{i}g_\mathrm{\bathw} &\mathrm{i}\omega_\mathrm{\bathw}-\frac{1}{2}(3\gamma^++\gamma^-) &0 &-\mathrm{i}g_\mathrm{\bathw}\\
0 &-\mathrm{i}g_\mathrm{\bathw} &0 &-\mathrm{i}\omega_\mathrm{\bathw}-\frac{1}{2}(3\gamma^++\gamma^-) &\mathrm{i}g_\mathrm{\bathw}\\
\gamma^- &\gamma^- &-\mathrm{i}g_\mathrm{\bathw} &\mathrm{i}g_\mathrm{\bathw} &-2\gamma^+
\end{bmatrix}.
 \end{align}
 
The block $\mathcal{L}_{5\times 5}$ is thus specified by the four parameters $\{\omega_\bathw, g_\bathw, \gamma^+,\gamma^-\}$. Since the block $\mathcal{L}_{5\times 5}$ includes all the terms related to the heat current and the steady state of the zero eigenvalue, we analyze the block, varying the values of $\{\omega_\bathw, g_\bathw, \gamma^+,\gamma^-\}$.
 
 \subsection{$g_\bathc$-coupling system: $g_\bathc\neq 0 and g_\bathw = 0$}
\label{A1-2}
If we neglect the coupling $g_\bathw $ and keep only $g_\bathc$, the three blocks in Eq.~(\ref{Lblock}) read
\begin{align}
&\mathcal{L}_{2 \times 2} =\begin{bmatrix}
\mathrm{i} \omega_\bathh -\Gamma_h &-\mathrm{i} g_\bathc\\
-\mathrm{i} g_\bathc & \mathrm{i} \omega_\bathw-\Gamma_\bathw \\
\end{bmatrix},\\
&\mathcal{L}_{5\times 5}=\begin{bmatrix}
-\gamma_\bathh^- -\gamma_\bathc^- &\mathrm{i} g_\bathc &-\mathrm{i} g_\bathc &\gamma_\bathc^+ &\gamma_\bathh^+ \\
\mathrm{i} g_\bathc &\mathrm{i} \omega_\bathc - \Gamma_\bathc &0 &-\mathrm{i} g_\bathc &0 \\
-\mathrm{i} g_\bathc &0 &- \mathrm{i} \omega_\bathc-\Gamma_\bathc &\mathrm{i} g_\bathc &0 \\
\gamma_\bathc^- &-\mathrm{i} g_\bathc &\mathrm{i} g_\bathc &-\gamma_\bathc^+-\gamma_\bathw^- &\gamma_\bathw^+ \\
\gamma_\bathh^- &0 &0 &\gamma_\bathw^- &-\gamma_\bathw^+ -\gamma_\bathh^+
\end{bmatrix}.
\end{align}
The basis of $\mathcal{L}_{2 \times 2}$ is $\begin{bmatrix}
 \rho_\textrm{gh} &\rho_\textrm{ch}
  \end{bmatrix}^T$, 
 that of $\mathcal{L}_{2 \times 2}^*$ is $\begin{bmatrix} 
 \rho_\textrm{hg} &\rho_\textrm{hc}
 \end{bmatrix}^T$, and that of $\mathcal{L}_{5\times 5}$ is $\begin{bmatrix} \rho_\textrm{gg} &\rho_\textrm{gc} &\rho_\textrm{cg} &\rho_\textrm{cc} &\rho_\textrm{hh} \end{bmatrix}^T$.
 If we set $\gamma^+=\gamma_\bathc^+=\gamma_\bathh^+=\gamma_\bathw^+$ and $\gamma^-=\gamma_\bathc^-=\gamma_\bathh^-=\gamma_\bathw^-$, we have
 \begin{align}
 &\mathcal{L}_{2 \times 2} =\begin{bmatrix}
\mathrm{i} \omega_\bathh -(\gamma^+ +\gamma^-) &-\mathrm{i} g_\bathc\\
-\mathrm{i} g_\bathc & \mathrm{i} \omega_\bathw-\frac{1}{2}(3\gamma^+ +\gamma^-) \\
\end{bmatrix},\\
 &\mathcal{L}_{5\times 5}=\begin{bmatrix}
-2\gamma^-  &\mathrm{i} g_\bathc &-\mathrm{i} g_\bathc &\gamma^+ &\gamma^+ \\
\mathrm{i} g_\bathc &\mathrm{i} \omega_\bathc - \frac{1}{2}(3\gamma^-+\gamma^+) &0 &-\mathrm{i} g_\bathc &0 \\
-\mathrm{i} g_\bathc &0 &- \mathrm{i} \omega_\bathc-\frac{1}{2}(3\gamma^-+\gamma^+) &\mathrm{i} g_\bathc &0 \\
\gamma^- &-\mathrm{i} g_\bathc &\mathrm{i} g_\bathc &-\gamma^+ -\gamma^- &\gamma^+ \\
\gamma^- &0 &0 &\gamma^- &-2\gamma^+
\end{bmatrix}.
 \end{align}
Similarly to the previous case, the block $\mathcal{L}_{5\times 5}$ is specified by four parameters $\{\omega_\bathc, g_\bathc, \gamma^+,\gamma^-\}$. Since the block $\mathcal{L}_{5\times 5}$ includes all the terms related to the heat current and the steady state of the zero eigenvalue, we analyze the block, varying the values of $\{\omega_\bathc, g_\bathc \gamma^+,\gamma^-\}$.

\section{Eigenvalue Distribution $\&$ Derivation of Third-Order LEPs}
\label{A2}

In this appendix, we derive the roots of the cubic polynomial equation $F_3(\lambda)$ in Eq.~(\ref{L5Det}) of the $g_\bathc$- and $g_\bathw$-coupling systems. We can write down all the roots of any third-degree polynomial equation $F_3(x)=ax^3+bx^2+cx+d=0$ in the general form:
\begin{align}
&x = -\frac{1}{3a}\left[b+\xi^{(v-2)} \mathcal{C} + \frac{\Delta_0}{\xi^{(v-2)} \mathcal{C}}\right] \text{   for } v=0,1,2 ,
\end{align}
where
\begin{align}
\label{xi}
&\xi = \frac{-1-i\sqrt{3}}{2},\\
&\mathcal{C} = \sqrt[3]{\frac{\Delta_1\pm \sqrt{\Delta_1^2 - 4\Delta_0^3}}{2}},\\
\label{Poly3-D0}
&\Delta_0 = b^2-3ac,\\
\label{Poly3-D1}
&\Delta_1 = 2b^3-9abc+27a^2d.
\end{align}
The discriminant of the cubic polynomial equation is given by
\begin{align}
\label{Poly3-D}
&\Delta = 18abcd-4b^3 d+b^2 c^2-4a c^3-27a^2 d^2.
\end{align}
We will specify the coefficients $a,b,c,d,$ below.
We can use the discriminant $\Delta$ and the quantity $\Delta_0$ to analyze the distribution of the roots as follows:
\begin{enumerate}[(a)]
\item $\Delta>0$: three distinct real roots;
\item $\Delta<0$: one real root and two non-real complex conjugate roots;
\item $\Delta=0$ but $\Delta_0\neq0$: a double root $x_2 = x_3 = (9ad-bc)/[2(b^2-3ac)]$ and a simple root $x_1 = (4abc-9a^2d-b^3)/[a(b^2-3ac)]$;
\item $\Delta=0$ and $\Delta_0=0$: a triple root $x_1 = x_2 = x_3=b/(3a)$.
\end{enumerate}

Together with the zero eigenvalue $\lambda_0$ and an independent real eigenvalue $\lambda_1$ in Eq.~(\ref{L5Det}), as in the main text, the distributions of eigenvalues for the $g_\bathw$- and $g_\bathc$-coupling systems is classified into two situations away from the LEPs: (a) zero eigenvalue and four distinct real eigenvalues, as exemplified Fig~\ref{nonEP1}; (b) zero-eigenvalue, two distinct real eigenvalues, and a pair of complex conjugate eigenvalues, as exemplified Fig~\ref{nonEP2}. 
We can generally write down the entire eigenvalues of the $5\times 5$ Liouvillian superoperator block $\hat{\mathcal{L}}_{5\times 5}$ away from the LEPs in the following forms: 
\begin{align}
&\lambda_0 = 0;\\
&\lambda_1 <0;\\
&\lambda_k = -\frac{1}{3a}\left [b+\xi^{(k-2)} \mathcal{C} + \frac{\Delta_0}{\xi^{(k-2)} \mathcal{C}}\right ] \text{    for } k=2,3,4,
\end{align}
where $\xi$, $\mathcal{C}$, $\Delta_0$ and $\Delta_1$ are as in Eqs.~(\ref{xi})--(\ref{Poly3-D1}).

At the second-order LEP, the pair of complex conjugate roots $\lambda_{3}$ and $\lambda_{4}$ coalesce to one real double root, as exemplified in Fig.~\ref{EP2}. At the third-order exceptional point, the pair of the complex conjugate roots $\lambda_{3}$ and $\lambda_{4}$ and one real root $\lambda_2$ coalesce to one real triple root, as exemplified in Fig.~\ref{EP3}.

In each of the $g_\bathw$- and $g_\bathc$-coupling systems, the coefficients are given by the following.
\begin{enumerate}
\item[(i)] $g_\bathc=0 \text{ and }  g_\bathw \neq 0$:
\begin{align}
&a = 4,\\
&b = 8\gamma^- + 20\gamma^+,\\
&c = 16 g_\bathw^2 + 5{\gamma^-}^2 + 26\gamma^-\gamma^+ +33{\gamma^+}^2+4\omega_\bathw^2,\\
&d = 8g_\bathw^2\gamma^- + {\gamma^-}^3+24 g_\bathw^2 \gamma^+ +8{\gamma^-}^2 \gamma^+ + 21\gamma^- {\gamma^+}^2 + 18 {\gamma^+}^3+4\gamma^- \omega_\bathw^2 + 8\gamma^+ \omega_\bathw^2.
\end{align}
Then we obtain the simple real root $\lambda_1$ and the triple root $\lambda^{(3)}$ as follows:
\begin{align}
&\lambda_1 = -(2\gamma^-+\gamma^+),\\
&\lambda^{(3)} =  \gamma^- -10\sqrt{3}\omega_\bathw.
\end{align}

\item[(ii)] $g_\bathc\neq 0 \text{ and } g_\bathw = 0$:
\begin{align}
&a = 4,\\
&b = 8\gamma^+ + 20\gamma^-,\\
&c = 16 g_\bathc^2 + 5{\gamma^+}^2 + 26\gamma^-\gamma^+ +33{\gamma^-}^2+4\omega_\bathc^2,\\
&d = 8g_\bathc^2\gamma^+ + {\gamma^+}^3+24 g_\bathc^2 \gamma^- +8{\gamma^+}^2 \gamma^- + 21\gamma^+ {\gamma^-}^2 + 18 {\gamma^-}^3+4\gamma^+ \omega_\bathc^2 + 8\gamma^- \omega_\bathc^2.
\end{align}
Then we obtain the simple real root $\lambda_1$ and the triple root $\lambda^{(3)}$ as follows:
\begin{align}
&\lambda_1 = -(2\gamma^++\gamma^-),\\
&\lambda^{(3)} =  \gamma^+ -10\sqrt{3}\omega_\bathc.
\end{align}
\end{enumerate}

\begin{figure}
		\centering
		\includegraphics[width=0.45\textwidth]{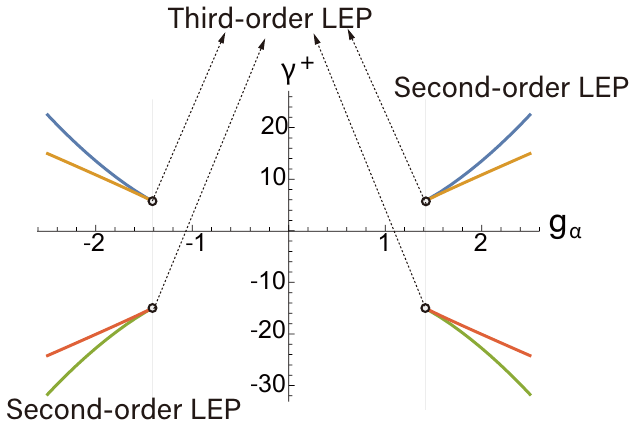}
		\caption{Second-order and third-order EPs. If we specify the parameter $\omega_\alpha$ as the energy unit with $\omega_\alpha=1$, the lines show the parameters $\gamma^+$ depending on the other parameter $g_\alpha$ at the second-order exceptional points. The crossing points of these lines are the third-order exceptional points.}
	\label{EP}
\end{figure}

Although the coefficients of the general cubic polynomial equations in the $g_\bathw$- and $g_\bathc$-coupling systems are different from each other, the relations (\ref{Poly3-D0}) and (\ref{Poly3-D1}) given by $\Delta_0=0$ and $\Delta_1=0$ at the triple root are similar to each other. There are four possibilities of $\gamma^+$ depending on $\gamma^-$, $\omega_\alpha$, and $g_\alpha$ when $\Delta_0=0$ in the $g_\alpha$-coupling system $(\alpha=w,c)$:
\begin{align}
\gamma^+ = -\gamma^- \pm 2 \sqrt{\frac{2 g_\alpha^4 + 10 g_\alpha^2 \omega_\alpha^2 -\omega_\alpha^4 \pm 2 \sqrt{g_\alpha^2 (g_\alpha^2-2\omega_\alpha^2)^3}}{\omega_\alpha^2}}.
\end{align}
When $\Delta_0=0$ is satisfied, there are double roots in the system. These double roots cross with each other on several surfaces in the four-dimensional space constructed by the four parameters ${\gamma^+,\gamma^-,\omega_\alpha,g_\alpha}$ for the $g_\alpha$-coupling system $(\alpha=\bathw,\bathc)$. We specify two parameters with constant values and vary the other two parameters. These second-order LEPs are lines in the two-dimensional parameter spaces, and overlap at several points, as shown in Fig.~\ref{EP}. These crossing points are comprised of third-order LEPs. 

Therefore, at all third-order LEPs with the eigenvalue $\lambda^{(3)}$, there are two relations between the four parameters, and hence we can rewrite the system dynamics depending on two of the parameters by setting the two other parameters as in
\begin{align}
\gamma^+ = -\gamma^- \text{ and } g_\alpha = \pm \frac{i \omega_\alpha}{2},
\end{align}
or
\begin{align}
\gamma^+ = -\gamma^- \pm 6\sqrt{3}\omega_\alpha  \text{ and }  g_\alpha = \pm \sqrt{2} \omega_\alpha.
\end{align}
We require all the parameters to be real and positive quantities, so that there is only one possible choice of the relations among the parameters:
\begin{align}
\label{relationEP3}
\gamma^+ =-\gamma^- + 6\sqrt{3}\omega_\alpha  \text{ and }  g_\alpha = \sqrt{2} \omega_\alpha.
\end{align}
These are presented in Eqs.~(\ref{EP3Para1}) and (\ref{EP3Para2})

\section{Generalized Eigensystem}
\label{A3}

Different from the diabolic points (DPs), the eigenvectors at the EPs also coalesce. There are two methods of checking the eigenvectors at the EPs
\begin{enumerate}
\item[I.] If there is only one solution $\rho$ for $(\mathcal{L}-\lambda^{(n)})\rho=0$, there is no degenerate DPs. The eigenvalue $\lambda^{(n)}$ is the one at the $ n$th-order EP instead of DPs.
\item[II.] After the Jordan decomposition, we find an $n\times n$ Jordan block at an $n$th-order EP.
\end{enumerate}
Using the above methods, we confirm that the third-order LEP appears in both $g_\bathw$- and $g_\bathc$-coupling systems. 

In the $g_\bathw$-coupling system, we can obtain only one solution:
 \begin{align}
 \rho=\begin{bmatrix} \rho_\textrm{gg} &\rho_\textrm{cc} &\rho_\textrm{ch} &\rho_\textrm{hc} &\rho_\textrm{hh} \end{bmatrix}^T = \begin{bmatrix}0 &\rho_\textrm{cc} &-2i\sqrt{2}\rho_\textrm{cc}/(i+\sqrt{3}) &-\sqrt{2}(-i+\sqrt{3})/(i+\sqrt{3})\rho_\textrm{cc} &-\rho_\textrm{cc} \end{bmatrix}^T.
 \end{align}
The Jordan decomposition is
\begin{align}
\mathcal{J}_{EP}^{(3)}=\begin{bmatrix}
0 &0 &0 &0 &0\\
0 &\gamma^- -10\sqrt{3}\omega_\bathc &1 &0 &0\\
0 & 0 & \gamma^- -10\sqrt{3}\omega_\bathc &1 &0\\
0 & 0 & 0 &\gamma^- -10\sqrt{3}\omega_\bathc  &0\\
0 & 0 & 0 &0 &-\gamma^- -6\sqrt{3}\omega_\bathc
\end{bmatrix}.
\end{align}
In the $g_\bathc$-coupling system, we can obtain only one solution:
\begin{align}
\rho=\begin{bmatrix} \rho_\textrm{gg} &\rho_\textrm{gc} &\rho_\textrm{cg} &\rho_\textrm{cc} &\rho_\textrm{hh} \end{bmatrix}^T = \begin{bmatrix}\rho_\textrm{gg} &-2i\sqrt{2}/(i+\sqrt{3})\rho_\textrm{gg} &-\sqrt{2}(-i+\sqrt{3})/(i+\sqrt{3})\rho_\textrm{gg} &-\rho_\textrm{gg} &0\end{bmatrix}^T.
\end{align}
The Jordan decomposition is
\begin{align}
\mathcal{J}_{EP}^{(3)}=\begin{bmatrix}
0 &0 &0 &0 &0\\
0 &\gamma^- -12\sqrt{3}\omega_\bathc &0 &0 &0\\
0 & 0 & -\gamma^- -4\sqrt{3}\omega_\bathc &1 &0\\
0 & 0 & 0 &-\gamma^- -4\sqrt{3}\omega_\bathc &1\\
0 & 0 & 0 &0 &-\gamma^- -4\sqrt{3}\omega_\bathc
\end{bmatrix}.
\end{align}
Both $g_\bathw$- and $g_\bathc$-coupling systems can be decomposed into the Jordan forms with a $3\times 3$ non-diagonalized block. Since there is only one eigenvector for the $3\times 3$ block, for completeness, we introduce two pseudo-eigenvectors for generalizing the eigensystem.

The generalized eigensystems at exceptional points can be derived from the Jordan chain. For the $n$th-order EP with the eigenvalue $\lambda^{(n)}$, we can obtain the pseudo-left eigenvector $\sigma_{EP}^{(i)}$ and pseudo-right one $\rho_{EP}^{(j)}$, where $i$ and $j$ are integers from 1 to $n$:
\begin{align}
(\tilde{\mathcal{L}}-\lambda^{(n)})\rho_{EP}^{(1)} = 0,\\
(\tilde{\mathcal{L}}-\lambda^{(n)})\rho_{EP}^{(2)} = \rho_{EP}^{(1)},\\
(\tilde{\mathcal{L}}-\lambda^{(n)})\rho_{EP}^{(3)} = \rho_{EP}^{(2)},\\
...\\
(\tilde{\mathcal{L}}-\lambda^{(n)})\rho_{EP}^{(n)} = \rho_{EP}^{(n-1)},
\end{align}
and
\begin{align}
\sigma_{EP}^{(n)}(\tilde{\mathcal{L}}-\lambda^{(n)}) = 0,\\
\sigma_{EP}^{(n-1)}(\tilde{\mathcal{L}}-\lambda^{(n)}) = \sigma_{EP}^{(n)},\\
\sigma_{EP}^{(n-2)}(\tilde{\mathcal{L}}-\lambda^{(n)}) = \sigma_{EP}^{(n-1)},\\
...\\
\sigma_{EP}^{(1)}(\tilde{\mathcal{L}}-\lambda^{(n)}) = \sigma_{EP}^{(2)}.
\end{align}
Note that only $\sigma^{(3)}_{EP}$ and $\rho^{(1)}_{EP}$ are the true eigenvectors.

In the main text, since we only focus on the one-coupling system at the third-order LEPs, there are three steps for the derivations of the generalized eigensystem. The generalized eigensystem is comprised of three pairs of pseudo-left and pseudo-right eigenvectors that satisfy
\begin{align}
\sigma_{EP}^{(i)}\rho_{EP}^{(j)}=\delta_{i,j} \text{  for } i,j=1,2,3.
\end{align}
Therefore, we can rewrite the system state at the third-order LEP as Eq.~(\ref{rhoEP3}) in Sec.~\ref{Sec2-3}.

\section{Critical Damping: Comparison between third-order LEP and near-LEP}
\label{A4}

Based on the derivation of the general solutions of the cubic polynomial equations in Appendix~\ref{A2}, we find the eigenvalues near LEPs as the sum of the eigenvalue at the LEP and extra terms. We can thereby rewrite the ratio $\mathcal{R}_\textrm{s}(t)$ in Eq.~(\ref{Rst}) as follows:
\begin{align}
\mathcal{R}_\textrm{s}(t) = &\frac{|| c_1 e^{\lambda_1t} \rho_1 + \bigl[ (c_2+c_3 t + c_4 \frac{t^2}{2}) \rho_{EP}^{(1)} +(c_3+c_4 t ) \rho_{EP}^{(2)} +c_4\rho_{EP}^{(3)} \bigr]e^{\lambda^{(3)}t}
||_1}{|| \tilde{c}_1 e^{\tilde{\lambda}_1t} \tilde{\rho}_1+ \tilde{c}_2 e^{\tilde{\lambda}_2t} \tilde{\rho}_2 +\tilde{c}_3 e^{\tilde{\lambda}_3t} \tilde{\rho}_3+ \tilde{c}_4 e^{\tilde{\lambda}_4t} \tilde{\rho}_4 ||_1}\\
=&\frac{|| c_1 e^{(\lambda_1-\lambda^{(3)})t} \rho_1 + (c_2+c_3 t + c_4 \frac{t^2}{2}) \rho_{EP}^{(1)} +(c_3+c_4 t ) \rho_{EP}^{(2)} +c_4\rho_{EP}^{(3)} 
||_1}{|| \tilde{c}_1 e^{(\tilde{\lambda}_1-\lambda^{(3)})t} + \tilde{c}_2 e^{(\tilde{\lambda}_2-\lambda^{(3)})t} +\tilde{c}_3 e^{(\tilde{\lambda}_3-\lambda^{(3)})t} + \tilde{c}_4 e^{(\tilde{\lambda}_4-\lambda^{(3)})t} ||_1}\\
\label{RsFinal}
=&\frac{|| c_1 e^{(\lambda_1-\lambda^{(3)})t} \rho_1 + (c_2+c_3 t + c_4 \frac{t^2}{2}) \rho_{EP}^{(1)} +(c_3+c_4 t ) \rho_{EP}^{(2)} +c_4\rho_{EP}^{(3)} 
||_1}{|| \tilde{c}_1 e^{(\tilde{\lambda}_1-\lambda^{(3)})t} + \tilde{c}'_2 e^{-\frac{1}{3 a}(\mathcal{C}+\frac{\Delta_0}{\mathcal{C}})t} +\tilde{c}'_3 e^{\frac{1}{6 a}((\mathcal{C}+\frac{\Delta_0}{\mathcal{C}})-i\sqrt{3}(\mathcal{C}-\frac{\Delta_0}{\mathcal{C}}))t} + \tilde{c}'_4 e^{\frac{1}{6 a}((\mathcal{C}+\frac{\Delta_0}{\mathcal{C}})+i\sqrt{3}(\mathcal{C}-\frac{\Delta_0}{\mathcal{C}}))t} ||_1},
\end{align}
where
\begin{align}
&\tilde{\lambda}_1 \approx \lambda_1,\\
&\tilde{\lambda}_k \approx \lambda^{(3)}-\frac{1}{3a}\Bigl[\xi^{(k-2)} \mathcal{C} + \frac{\Delta_0}{\xi^{(k-2)} \mathcal{C}}\Bigr], \text{   for } k=2,3,4.
\end{align}
This indicates that the eigenstate converges to the steady state faster when the eigenvalue is closer to the steady state's eigenvalue. In particular, if $\lambda^{(3)}$ is closer to the zero eigenvalue than $\lambda_1$ and $\tilde{\lambda}_1$, the third-order LEP $\lambda^{(3)}$ plays the dominant role. 

In the numerator of Eq.~(\ref{RsFinal}), the factor $\exp[(\lambda_1-\lambda^{(3)})t]$ decreases exponentially and converges to zero in the long-term evolution when $\lambda_1<\lambda^{(3)}$. Since both $\lambda_1$ and $\lambda^{(3)}$ are negative quantities for real positive parameters, the inequality $|\lambda^{(3)}|<|\lambda_1|$ means that the eigenvalue at the third-order LEP is closer to the zero eigenvalue. Therefore, the numerator of $\mathcal{R}_\textrm{s}(t)$ is mainly influenced by the terms related to $t$ and $t^2$ and grows as $\mathcal{O}(t^2)$. 

For the denominator, if $\mathcal{C}+\Delta_0/\mathcal{C}<0$, 
the denominator of $\mathcal{R}_\textrm{s}(t)$ is influenced by the terms related to $\exp\{{[1/(3a)]|\mathcal{C}+\Delta_0/\mathcal{C}|t}\}$ and grows as $\mathcal{O}[\exp(t)]$. Inversely, if $\mathcal{C}+\Delta_0/\mathcal{C}>0$, 
the denominator of $\mathcal{R}_\textrm{s}(t)$ is influenced by the terms that relate to $\exp\{{[1/(6a)]|\mathcal{C}+\Delta_0/\mathcal{C}|t}\}$ and grows as $\mathcal{O}[\exp(t)]$. Therefore, if the influence of the eigenvalue $\lambda_1$ is less than the other non-EP state, the denominator grows as $\mathcal{O}[\exp(t)]$.
 
Since the numerator grows as $\mathcal{O}(t^2)$ while the denominator grows as $\mathcal{O}(e^t)$, the system at the third-order LEP is closer to the steady state than the system at the non-LEP state. It is easy to confirm that the critical damping~(\ref{RsCD}) results and that the state at the EP converges to the steady state faster than at the non-EP state.

If we rewrite the circumstance $\lambda_1<\lambda^{(3)}<0$ in terms of $\gamma^-$ and $\omega_\alpha$, we can find the relation between these two parameters $\{\gamma^-$, $\omega_\alpha\}$ as in Eqs.~(\ref{gwCDcondition}) and (\ref{gcCDcondition}) for the system to achieve critical damping.
Similarly, since the heat currents $\dot{Q}_\alpha(t)$ are given by the weighted sum of the system states, under the same circumstances in Eqs.~(\ref{gwCDcondition}) and (\ref{gcCDcondition}), the heat currents can achieve the critical damping, given the ratio $\mathcal{R}_\bathc(t)$~(\ref{Rct}) and the condition~(\ref{RcCD}).

\section{Choice of Initial State}
\label{A5}

The improvement of performance happens under restricted conditions. We analyze the influence of the third-order LEP terms on the heat current in this section for choosing a proper initial state. 

We require the following two constraints for better performance:
\begin{enumerate}
\item[(I)] more heat transfer: $\dot{Q}_\bathc(t)>\dot{Q}_\bathc^\textrm{ss}>0$ and $\dot{Q}_\bathh(t)<\dot{Q}_\bathh^\textrm{ss}<0$;
\item[(II)] higher coefficient of performance (COP): $\eta_\textrm{inst}(t)=\dot{Q}_\bathc(t)/\dot{Q}_\bathw(t)>\dot{Q}_\bathc^\textrm{ss}/\dot{Q}_\bathw^\textrm{ss}>0$.
\end{enumerate}
In addition, based on the operation of a refrigerator, the requirements $\dot{Q}_\bathc(t)>0$, $\dot{Q}_\bathw(t)>0$ and $\dot{Q}_\bathh(t)<0$ should be satisfied. Therefore, we rewrite all the requirements as follows:
\begin{align}
\dot{Q}_\bathc(t)>\dot{Q}_\bathc^\textrm{ss}>0\text{, }\dot{Q}_\bathh(t)<\dot{Q}_\bathh^\textrm{ss}<0\text{ and }0<\dot{Q}_\bathw(t)<\dot{Q}_\bathw^\textrm{ss}<0.
\end{align}

We only consider the systems under the critical-damping conditions~(\ref{gwCDcondition}) and (\ref{gcCDcondition}). We compare the initial heat current $\dot{Q}_\alpha^i=\dot{Q}_\alpha(0)$ to its equilibrium heat $\dot{Q}_\alpha^\textrm{ss}$. Note that the equilibrium heat current $\dot{Q}_\alpha^\textrm{ss}$ is not influenced by the initial state $\rho(0)$ in Eq.~(\ref{rho0}).

\subsection{$g_\bathw$-coupling system}
\label{A5-1}

\subsubsection{Influence of Initial State}
\begin{figure*}
	\centering
	\begin{subfigure}{0.4\textwidth}
		\includegraphics[width=1\textwidth]{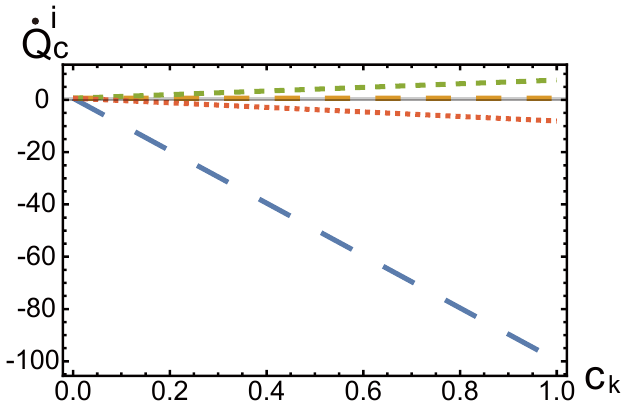}
		\caption{$\dot{Q}_\bathc^i(c_k)$}	
		\label{gwQcick}
	\end{subfigure}
	\begin{subfigure}{0.4\textwidth}
		\includegraphics[width=1\textwidth]{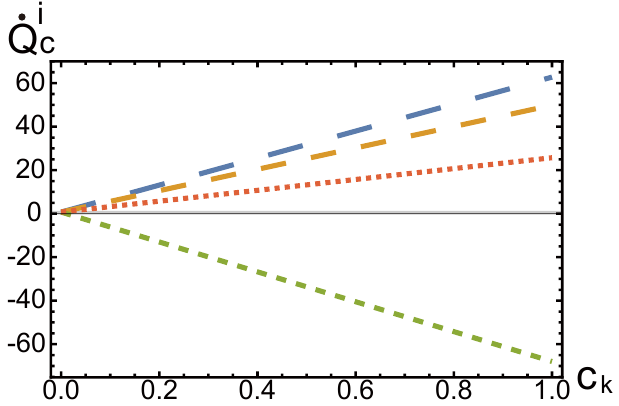}
		\caption{$\dot{Q}_\bathw^i(c_k)$}
		\label{gwQwick}
	\end{subfigure}\par
	\begin{subfigure}{0.4\textwidth}
		\centering
		\includegraphics[width=1\textwidth]{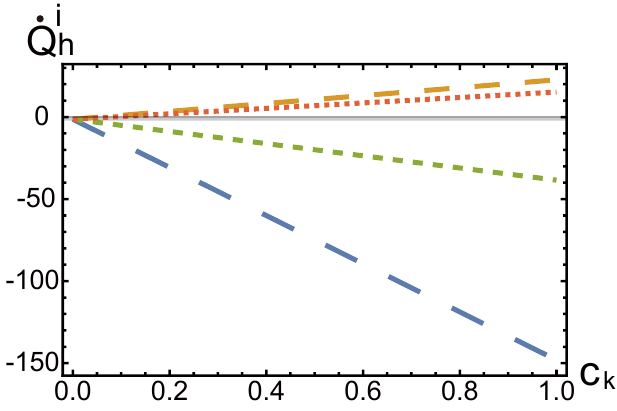}
		\caption{$\dot{Q}_\bathh^i(c_k)$}
		\label{gwQhick}
	\end{subfigure}
	\begin{subfigure}{0.4\textwidth}
		\centering
		\includegraphics[width=0.5\textwidth]{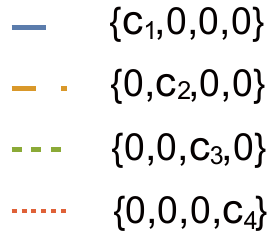}
		\caption*{ }
	\end{subfigure}
	\caption{Dependence of the initial non-equilibrium heat current $\dot{Q}_\alpha^i$ on each coefficient $c_k$ ($k=1,2,3,4$) for $g_\bathw$-coupling system. We analyze the influence of each coefficient $c_k$ by setting the other three coefficients to zero. Parameters: dissipation rate $\gamma^-=7$, energy levels $\omega_\bathw=\omega_\bathc=1$ are equal to energy unit.}
	\label{gwQick}
\end{figure*}

As shown in Fig.~\ref{gwQick}, for the $g_\bathw$-coupling system, as we increase each coefficient $c_k$, the initial non-equilibrium heat current $\dot{Q}_\bathc^i$ from the cold bath increases for $c_{3}$ but decreases for $c_1$ and $c_4$. The coefficient $c_2$ does not change $\dot{Q}_\bathc^i$. The heat current $\dot{Q}_\bathw^i$ from the work bath increases for $c_1$, $c_2$ and $c_4$ but decreases for $c_{3}$. The heat current $\dot{Q}_\bathh^i$ increases for $c_2$ and $c_4$ but decreases for the coefficients $c_1$ and $c_3$. 

For achieving the two constraints of better performance, we try to increase $\dot{Q}_\bathc^i$ but decrease $\dot{Q}_\bathw^i$ and $\dot{Q}_\bathh^i$. The coefficient $c_3$ is dominant for achieving not only more heat transfer but also higher COP, since it increases $\dot{Q}_\bathc^i$ and decreases $\dot{Q}_\bathw^i$ and $\dot{Q}_\bathh^i$. The coefficient $c_2$ decreases $\dot{Q}_\bathw^i$ and does not change $\dot{Q}_\bathc^i$, and hence it also improves the $g_\bathw$-coupling system's performance with higher COP. The speed at which the coefficient $c_2$ increases the heat current $\dot{Q}_\bathh^i$ is slower than the coefficient $c_3$ decreases $\dot{Q}_\bathh^i$, and thereby the heat transfer of $g_\bathw$-coupling system is increased by increasing $c_2$ and $c_3$ at the same time.


\subsubsection{Optimal Performance}

For $\dot{Q}_\bathw=0$, we would achieve the highest heat absorption from the cold bath, and the COP diverges. Since the infinite COP is only caused by the zero denominator, instead of searching for the highest COP in the main sections, we rather focus on the phenomena that the third-order LEP state improves the system's COP and the optimal initial heat current $\dot{Q}_\bathc^i$ from the cold bath for the quantum absorption refrigerators.

\begin{figure}
	\centering
	\includegraphics[width=0.5\textwidth]{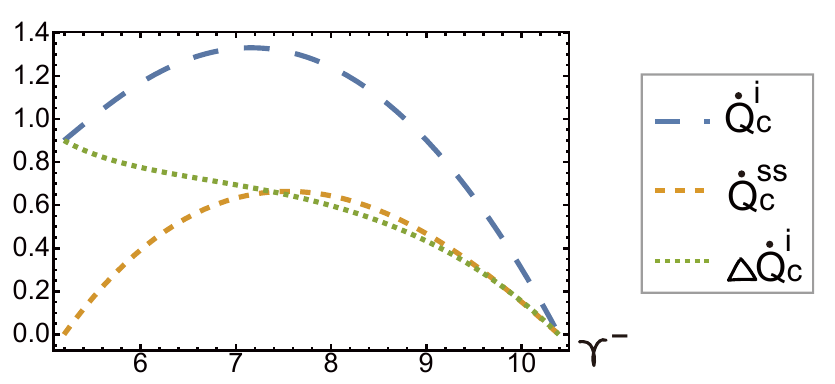}
	\caption{Dependence of the initial non-equilibrium heat current $\dot{Q}_\bathc^i$, equilibrium heat currnt $\dot{Q}_\bathc^\textrm{ss}$ and their difference $\Delta \dot{Q}_\bathc^i=\dot{Q}_\bathc^i-\dot{Q}_\bathc^\textrm{ss}$ on the dissipation rate $\gamma^-$ for the $g_\bathw$-coupling system with coefficient $\{c_1,c_2,c_3,c_4\}=\{0,0,0.1,0.1\}$, energy levels $\omega_\bathw=\omega_\bathc=1$ are equal to energy unit.}
	\label{gwQcigamma}
\end{figure}

Different from the coefficients of the non-equilibrium terms in the initial state, the dissipation rate $\gamma^-$ influences not only the non-equilibrium heat currents but also the equilibrium heat currents.  As shown in Fig.~\ref{gwQcigamma}, the dissipation rate $\gamma^-$ first increases the equilibrium heat $\dot{Q}_\bathc^\textrm{ss}$ and the initial non-equilibrium heat current $\dot{Q}_\bathc^i$ and then decreases them after the peak values. The difference $\Delta \dot{Q}_\bathc^i$ depends on the dissipation rate $\gamma^-$ and the coefficients $c_1$ through $c_4$. For a small value of $\gamma^-$, the influence of the coefficients on $\Delta \dot{Q}_\bathc^i$ is significant. As shown in Fig.~\ref{gwQcigamma}, the improvement of the heat current $\Delta \dot{Q}_\bathc^i$ becomes weaker when $\gamma^-$ increases.

Since the coefficient $c_3$ plays a significant role in increasing the heat current $\dot{Q}_\bathc^i$, the peak value of the non-equilibrium heat current $\dot{Q}_\bathc^i$ in Fig.~\ref{gwQcigamma} increases when $c_3$ increases. The equilibrium heat current does not depend on the coefficients of the non-equilibrium terms, and therefore the difference $\Delta \dot{Q}_\bathc^i$ increases if the non-equilibrium heat current $\dot{Q}_\bathc^i$ increases with the same dissipation rate $\gamma^-$. If we increase the coefficient $c_3$, the optimal heat current $\dot{Q}_\bathc^i$ and the heat difference $\Delta \dot{Q}_\bathc^i$ become larger. 

Based on the influence of each parameter in Figs.~\ref{gwQick} and \ref{gwQcigamma}, we find that a larger value of the non-equilibrium heat current $\dot{Q}_\bathc^i$ should be achieved by greater $c_3$ and smaller $c_1$ and $c_4$ with a proper $\gamma^-$. However, for a large value of $c_3$, the heat current $Q_\bathw^i$ becomes too small to be positive, which means that the $g_\bathw$-coupling system cannot run as a refrigerator in the non-equilibrium process. We only look for the optimal non-equilibrium heat current when the system satisfies the two constraints of better performance as a refrigerator. Since the coefficient $c_2$ increases the heat current $\dot{Q}_\bathw^i$ without changing $\dot{Q}_\bathc^i$, we can increase it to make the system still run as a refrigerator.

If we fix the value of $c_3$, we can always search for the optimal non-equilibrium heat current $\dot{Q}_\bathc^i$ by adjusting the other parameters $\{\gamma^-,c_1,c_2,c_4\}$. For example, if we set $c_3=0.1$, we numerically find the highest heat current $\dot{Q}_\bathc^i=1.330$ for $\{\gamma^-,c_1,c_2,c_4\}=\{7.159, 0, 0.125, 0\}$. We chose this set of parameters in the main sections.

\subsection{$g_\bathc$-coupling system}
\label{A5-2}

\subsubsection{Influence of Initial state}
\begin{figure*}
	\centering
	\begin{subfigure}{0.4\textwidth}
		\includegraphics[width=1\textwidth]{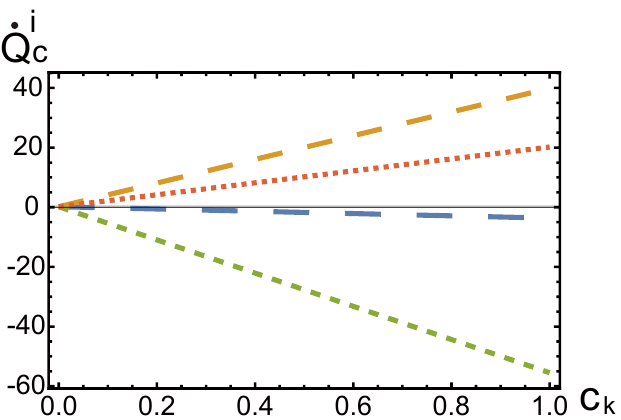}
		\caption{$\dot{Q}_\bathc^i(c_k)$}	
		\label{gcQcick}
	\end{subfigure}
	\begin{subfigure}{0.4\textwidth}
		\includegraphics[width=1\textwidth]{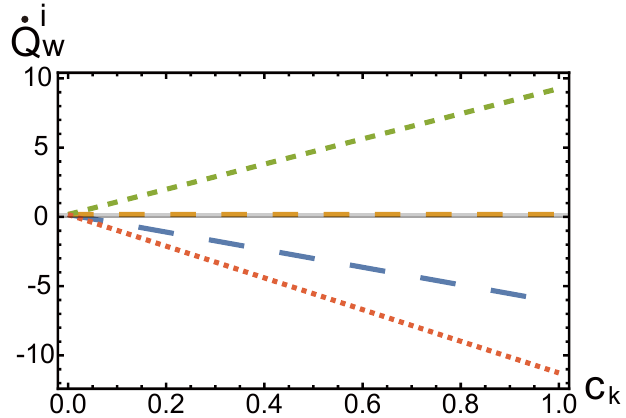}
		\caption{$\dot{Q}_\bathw^i(c_k)$}
		\label{gcQwick}
	\end{subfigure}\par
	\begin{subfigure}{0.4\textwidth}
		\centering
		\includegraphics[width=1\textwidth]{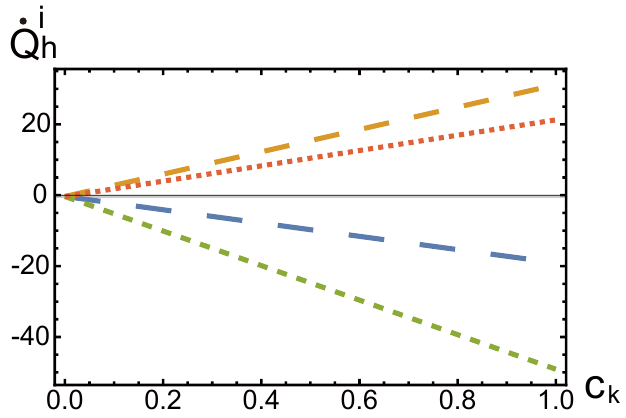}
		\caption{$\dot{Q}_\bathh^i(c_k)$}
		\label{gcQhick}
	\end{subfigure}
	\begin{subfigure}{0.4\textwidth}
		\centering
		\includegraphics[width=0.5\textwidth]{./Figs/ckLegend.pdf}
		\caption*{ }
	\end{subfigure}
	\caption{Dependence of the initial non-equilibrium heat current $\dot{Q}_\alpha^i$ on each coefficient $c_k$ ($k=1,2,3,4$) for $g_\bathc$-coupling system. We analyze the influence of each coefficient $c_k$ by setting the other three coefficients to zero. Parameters: dissipation rate $\gamma^-=5.5$, energy levels $\omega_\bathw=\omega_\bathc=1$ are equal to the energy unit.}
	\label{gcQick}
\end{figure*}

As shown in Fig.~\ref{gcQick}, for the $g_\bathc$-coupling system, as we increase each coefficient $c_k$, 
the initial non-equilibrium heat current $\dot{Q}_\bathc^i$ from the cold bath increases for $c_2$ and $c_4$ but decreases for the coefficients $c_1$ and $c_3$. The heat current $\dot{Q}_\bathw^i$ from the work bath increases for $c_3$ but decreases for $c_1$ and $c_4$. The coefficient $c_2$ does not influence the heat current $\dot{Q}_\bathw^i$. The heat current $\dot{Q}_\bathh^i$ increases for $c_2$ and $c_4$ but decreases for $c_1$ and $c_3$. 

We only look for the situations where the $g_\bathc$-coupling system can satisfy the two better-performance constraints and still run as a refrigerator. In trying to find the system to absorb more heat from the bath $\bathc$ and release more heat to the bath $\bathh$ with less energy cost from $\bathw$, the best choice is to keep $c_3$ and $c_4$ small and increase $c_1$ and $c_2$. As shown in Fig.~\ref{gcQick}, the coefficient $c_2$ increases $\dot{Q}_\bathc^i$ faster than $\dot{Q}_\bathh^i$, and $c_1$ decreases $\dot{Q}_\bathc^i$ slower than $\dot{Q}_\bathh^i$. Therefore, by increasing $c_1$ and $c_2$ at the same time, more heat transfer is achieved. In addition, since $c_1$ decreases $\dot{Q}_\bathw^i$, less energy cost from the work bath $\bathw$ is achieved. 

\subsubsection{Optimal Performance}

For $\dot{Q}_\bathw=0$, we would achieve the highest heat absorption from the cold bath, and the COP diverges. Since the infinite COP is only caused by the zero denominator, instead of searching for the highest COP in the main sections, we rather focus on the phenomena that the third-order LEP state improves the system's COP and the optimal initial heat current $\dot{Q}_\bathc^i$ from the cold bath for the quantum absorption refrigerators.

\begin{figure}
	\centering
	\begin{subfigure}{0.4\textwidth}
		\centering
		\includegraphics[width=1\textwidth]{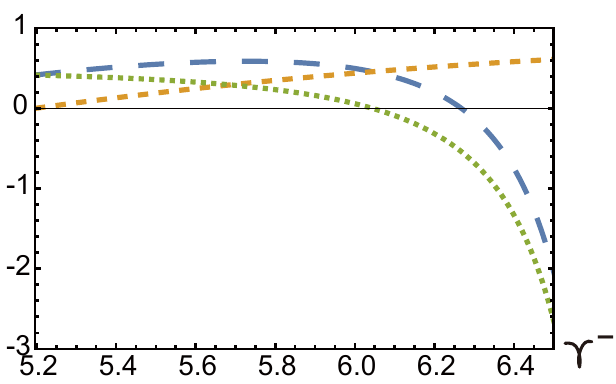}
		\caption{$\{c_1,c_2,c_3,c_4\}=\{0.01,0.01,0,0\}$}
		\label{gcQcigammaSmallc1c3}
	\end{subfigure}	
	\begin{subfigure}{0.4\textwidth}
		\centering
		\includegraphics[width=1\textwidth]{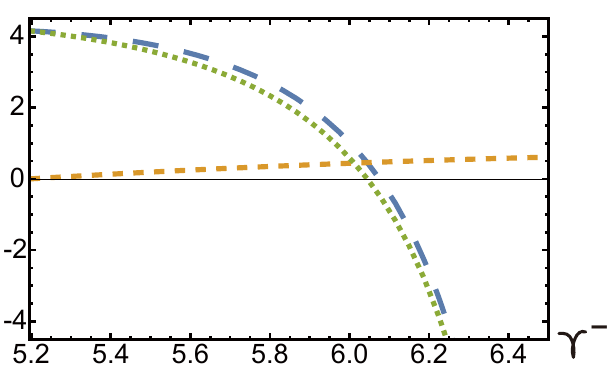}
		\caption{$\{c_1,c_2,c_3,c_4\}=\{0.1,0.1,0,0\}$}
		\label{gcQcigammaBigc1c3}
	\end{subfigure}	
	\begin{subfigure}{0.1\textwidth}
		\centering
		\includegraphics[width=1\textwidth]{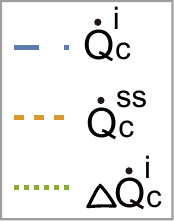}
		\caption*{ }
	\end{subfigure}	
		\caption{Dependence of the initial non-equilibrium heat current $\dot{Q}_\bathc^i$, equilibrium heat currnt $\dot{Q}_\bathc^\textrm{ss}$ and their difference $\Delta \dot{Q}_\bathc^i=\dot{Q}_\bathc^i-\dot{Q}_\bathc^\textrm{ss}$ on the dissipation rate $\gamma^-$ for the $g_\bathc$-coupling system with coefficients (a)~$\{c_1,c_2,c_3,c_4\}=\{0.01,0.01,0,0\}$ and (b)~$\{c_1,c_2,c_3,c_4\}=\{0.1,0.1,0,0\}$, energy levels $\omega_\bathw=\omega_\bathc=1$ are equal to energy unit.}
	\label{gcQcigamma}
\end{figure}

The dissipation rate $\gamma^-$ influences the non-equilibrium and equilibrium heat currents. We focus on $\dot{Q}_\bathc^i$ for searching the optimal heat absorption. As shown in Fig.~\ref{gcQcigamma}, the dissipation rate $\gamma^-$ increases the equilibrium heat $\dot{Q}_\bathc^\textrm{ss}$ linearly. With a small proportion of the third-order LEP state, the initial heat current $\dot{Q}_\bathc^i$ firstly increases and then decreases after a peak value as shown in Fig.~\ref{gcQcigammaSmallc1c3}. 

As shown in Fig.~\ref{gcQcigammaBigc1c3}, when the proportion of the third-order LEP state becomes large, the dissipation rate $\gamma^-$ decreases the initial non-equilibrium heat current $\dot{Q}_\bathc^i$. The decreasing speed becomes rapid when we increase the value of $\gamma^-$.
A small value of $\gamma^-$ does not decrease the initial heat current $\dot{Q}_\bathc^i$ much. Compared to the influence caused by the non-equilibrium state, the influence of $\gamma^-$ on the equilibrium heat current is quite small. With a small dissipation rate, the non-equilibrium heat current $\dot{Q}_\bathc^i$ is always larger than the equilibrium one by introducing the term $\rho_\textrm{EP}^{(1)}$ with the coefficient $c_2$ into the initial state. 

Figure~\ref{gcQick} shows that the coefficient $c_3$ decreases $\dot{Q}_\bathc^i$ much faster than $\dot{Q}_\bathh^i$. When we increase $c_3$, the heat current $\dot{Q}_\bathc^i$ decreases faster than $\dot{Q}_\bathh^i$, leading to less heat transfer. Similarly, the coefficient $c_1$ increases $\dot{Q}_\bathc^i$ slower than $\dot{Q}_\bathh^i$. When we increase $c_3$, the heat current $\dot{Q}_\bathc^i$ increases more slowly than $\dot{Q}_\bathh^i$, leading to less heat transfer.
Therefore, for the optimal heat current of the refrigerator, we should increase $c_1$ and $c_2$ but decrease $c_3$ and $c_4$ with a small dissipation rate $\gamma^-$.

The optimal heat current always increases when we increase $c_2$ with proper parameters $\{\gamma^-,c_1, c_3, c_4\}$. We can find the maximum heat current $\dot{Q}_\bathc^i$ by fixing the value of $c_2$. For example, if we set $c_2=0.1$, we numerically calculate the optimal heat current $1.93$ for $\{\gamma^-,c_1,c_3,c_4\}=\{5.857, 0.127,0, 0\}$.

\bibliographystyle{plainnat}
\bibliography{reference}

\begin{thebibliography}{70}
\providecommand{\natexlab}[1]{#1}
\providecommand{\url}[1]{\texttt{#1}}
\expandafter\ifx\csname urlstyle\endcsname\relax
  \providecommand{\doi}[1]{doi: #1}\else
  \providecommand{\doi}{doi: \begingroup \urlstyle{rm}\Url}\fi

\bibitem[Alicki(1979)]{RAlicki_1979}
R~Alicki.
\newblock {The quantum open system as a model of the heat engine}.
\newblock \emph{Journal of Physics A: Mathematical and General}, 12\penalty0
  (5):\penalty0 L103, may 1979.
\newblock \doi{10.1088/0305-4470/12/5/007}.
\newblock URL \url{https://dx.doi.org/10.1088/0305-4470/12/5/007}.

\bibitem[Angel~Rivas(2011)]{978-3-642-23354-8}
Susana F.~Huelga Angel~Rivas.
\newblock \emph{{Open Quantum Systems: An Introduction}}.
\newblock SpringerBriefs in Physics. Springer Berlin, Heidelberg, 10 2011.
\newblock ISBN 978-3-642-23354-8.
\newblock \doi{https://doi.org/10.1007/978-3-642-23354-8}.
\newblock URL \url{https://link.springer.com/book/10.1007/978-3-642-23354-8}.

\bibitem[Bergholtz et~al.(2021)Bergholtz, Budich, and
  Kunst]{RevModPhys.93.015005}
Emil~J. Bergholtz, Jan~Carl Budich, and Flore~K. Kunst.
\newblock {Exceptional topology of non-Hermitian systems}.
\newblock \emph{Rev. Mod. Phys.}, 93:\penalty0 015005, Feb 2021.
\newblock \doi{10.1103/RevModPhys.93.015005}.
\newblock URL \url{https://link.aps.org/doi/10.1103/RevModPhys.93.015005}.

\bibitem[Berry(2004)]{berry2004physics}
Michael~V Berry.
\newblock {Physics of nonhermitian degeneracies}.
\newblock \emph{Czechoslovak journal of physics}, 54\penalty0 (10):\penalty0
  1039--1047, 2004.

\bibitem[Bhandari and Jordan(2021)]{PhysRevB.104.075442}
Bibek Bhandari and Andrew~N. Jordan.
\newblock {Minimal two-body quantum absorption refrigerator}.
\newblock \emph{Phys. Rev. B}, 104:\penalty0 075442, Aug 2021.
\newblock \doi{10.1103/PhysRevB.104.075442}.
\newblock URL \url{https://link.aps.org/doi/10.1103/PhysRevB.104.075442}.

\bibitem[Bhandari et~al.(2020)Bhandari, Alonso, Taddei, von Oppen, Fazio, and
  Arrachea]{PhysRevB.102.155407}
Bibek Bhandari, Pablo~Terr\'en Alonso, Fabio Taddei, Felix von Oppen, Rosario
  Fazio, and Liliana Arrachea.
\newblock {Geometric properties of adiabatic quantum thermal machines}.
\newblock \emph{Phys. Rev. B}, 102:\penalty0 155407, Oct 2020.
\newblock \doi{10.1103/PhysRevB.102.155407}.
\newblock URL \url{https://link.aps.org/doi/10.1103/PhysRevB.102.155407}.

\bibitem[Bhattacharjee and Dutta(2021)]{Bhattacharjee2021}
Sourav Bhattacharjee and Amit Dutta.
\newblock {Quantum thermal machines and batteries}.
\newblock \emph{The European Physical Journal B}, 94\penalty0 (12):\penalty0
  239, 2021.
\newblock \doi{10.1140/epjb/s10051-021-00235-3}.
\newblock URL \url{https://doi.org/10.1140/epjb/s10051-021-00235-3}.

\bibitem[Brask and Brunner(2015)]{PhysRevE.92.062101}
Jonatan~Bohr Brask and Nicolas Brunner.
\newblock {Small quantum absorption refrigerator in the transient regime: Time
  scales, enhanced cooling, and entanglement}.
\newblock \emph{Phys. Rev. E}, 92:\penalty0 062101, Dec 2015.
\newblock \doi{10.1103/PhysRevE.92.062101}.
\newblock URL \url{https://link.aps.org/doi/10.1103/PhysRevE.92.062101}.

\bibitem[Breuer(2007)]{PhysRevA.75.022103}
Heinz-Peter Breuer.
\newblock {Non-Markovian generalization of the Lindblad theory of open quantum
  systems}.
\newblock \emph{Phys. Rev. A}, 75:\penalty0 022103, Feb 2007.
\newblock \doi{10.1103/PhysRevA.75.022103}.
\newblock URL \url{https://link.aps.org/doi/10.1103/PhysRevA.75.022103}.

\bibitem[Breuer and Petruccione(2007)]{9780199213900}
Heinz-Peter Breuer and Francesco Petruccione.
\newblock \emph{{The Theory of Open Quantum Systems}}.
\newblock Oxford University Press, 01 2007.
\newblock ISBN 9780199213900.
\newblock \doi{10.1093/acprof:oso/9780199213900.001.0001}.
\newblock URL \url{https://doi.org/10.1093/acprof:oso/9780199213900.001.0001}.

\bibitem[Campaioli et~al.(2024)Campaioli, Cole, and
  Hapuarachchi]{PRXQuantum.5.020202}
Francesco Campaioli, Jared~H. Cole, and Harini Hapuarachchi.
\newblock {Quantum Master Equations: Tips and Tricks for Quantum Optics,
  Quantum Computing, and Beyond}.
\newblock \emph{PRX Quantum}, 5:\penalty0 020202, Jun 2024.
\newblock \doi{10.1103/PRXQuantum.5.020202}.
\newblock URL \url{https://link.aps.org/doi/10.1103/PRXQuantum.5.020202}.

\bibitem[Carmichael(2009)]{carmichael2009open}
Howard Carmichael.
\newblock \emph{{An open systems approach to quantum optics: lectures presented
  at the Universit{\'e} Libre de Bruxelles, October 28 to November 4, 1991}},
  volume~18.
\newblock Springer Science \& Business Media, 2009.

\bibitem[Caves(1982)]{caves1982quantum}
Carlton~M Caves.
\newblock {Quantum limits on noise in linear amplifiers}.
\newblock \emph{Physical Review D}, 26\penalty0 (8):\penalty0 1817, 1982.

\bibitem[Chru\'{s}ci\'{n}ski and Pascazio(2017)]{doi:10.1142/S1230161217400017}
Dariusz Chru\'{s}ci\'{n}ski and Saverio Pascazio.
\newblock {A Brief History of the GKLS Equation}.
\newblock \emph{Open Systems \& Information Dynamics}, 24\penalty0
  (03):\penalty0 1740001, 2017.
\newblock \doi{10.1142/S1230161217400017}.
\newblock URL \url{https://doi.org/10.1142/S1230161217400017}.

\bibitem[Correa et~al.(2013)Correa, Palao, Adesso, and
  Alonso]{PhysRevE.87.042131}
Luis~A. Correa, Jos\'e~P. Palao, Gerardo Adesso, and Daniel Alonso.
\newblock {Performance bound for quantum absorption refrigerators}.
\newblock \emph{Phys. Rev. E}, 87:\penalty0 042131, Apr 2013.
\newblock \doi{10.1103/PhysRevE.87.042131}.
\newblock URL \url{https://link.aps.org/doi/10.1103/PhysRevE.87.042131}.

\bibitem[Correa et~al.(2014)Correa, Palao, Alonso, and Adesso]{Correa2014}
Luis~A. Correa, Jos{\'e}P. Palao, Daniel Alonso, and Gerardo Adesso.
\newblock {Quantum-enhanced absorption refrigerators}.
\newblock \emph{Scientific Reports}, 4\penalty0 (1):\penalty0 3949, 2014.
\newblock \doi{10.1038/srep03949}.
\newblock URL \url{https://doi.org/10.1038/srep03949}.

\bibitem[Dann et~al.(2018)Dann, Levy, and Kosloff]{dann2018time}
Roie Dann, Amikam Levy, and Ronnie Kosloff.
\newblock {Time-dependent Markovian quantum master equation}.
\newblock \emph{Physical Review A}, 98\penalty0 (5):\penalty0 052129, 2018.

\bibitem[Das~Agarwal et~al.(2024)Das~Agarwal, Konar, Lakkaraju, and
  Sen(De)]{PhysRevA.110.012226}
Keshav Das~Agarwal, Tanoy~Kanti Konar, Leela Ganesh~Chandra Lakkaraju, and
  Aditi Sen(De).
\newblock {Detecting exceptional points through dynamics in non-Hermitian
  systems}.
\newblock \emph{Phys. Rev. A}, 110:\penalty0 012226, Jul 2024.
\newblock \doi{10.1103/PhysRevA.110.012226}.
\newblock URL \url{https://link.aps.org/doi/10.1103/PhysRevA.110.012226}.

\bibitem[Friedman and Segal(2019)]{PhysRevE.100.062112}
Hava~Meira Friedman and Dvira Segal.
\newblock {Cooling condition for multilevel quantum absorption refrigerators}.
\newblock \emph{Phys. Rev. E}, 100:\penalty0 062112, Dec 2019.
\newblock \doi{10.1103/PhysRevE.100.062112}.
\newblock URL \url{https://link.aps.org/doi/10.1103/PhysRevE.100.062112}.

\bibitem[Gao and Hatano(2024)]{PhysRevResearch.6.023172}
Jingyi Gao and Naomichi Hatano.
\newblock {Maximum power of coupled-qubit Otto engines}.
\newblock \emph{Phys. Rev. Res.}, 6:\penalty0 023172, May 2024.
\newblock \doi{10.1103/PhysRevResearch.6.023172}.
\newblock URL \url{https://link.aps.org/doi/10.1103/PhysRevResearch.6.023172}.

\bibitem[Gardiner and Zoller(2004)]{gardiner2004quantum}
C.~Gardiner and P.~Zoller.
\newblock \emph{{Quantum Noise: A Handbook of Markovian and Non-Markovian
  Quantum Stochastic Methods with Applications to Quantum Optics}}.
\newblock Springer Series in Synergetics. Springer, 2004.
\newblock ISBN 9783540223016.
\newblock URL \url{https://books.google.co.jp/books?id=a_xsT8oGhdgC}.

\bibitem[Gelbwaser-Klimovsky et~al.(2013)Gelbwaser-Klimovsky, Alicki, and
  Kurizki]{PhysRevE.87.012140}
D.~Gelbwaser-Klimovsky, R.~Alicki, and G.~Kurizki.
\newblock {Minimal universal quantum heat machine}.
\newblock \emph{Phys. Rev. E}, 87:\penalty0 012140, Jan 2013.
\newblock \doi{10.1103/PhysRevE.87.012140}.
\newblock URL \url{https://link.aps.org/doi/10.1103/PhysRevE.87.012140}.

\bibitem[Geusic et~al.(1967)Geusic, Schulz-DuBios, and
  Scovil]{geusic1967quantum}
JE~Geusic, EO~Schulz-DuBios, and HED Scovil.
\newblock {Quantum equivalent of the carnot cycle}.
\newblock \emph{Physical Review}, 156\penalty0 (2):\penalty0 343, 1967.

\bibitem[Geva and Kosloff(1996)]{geva1996quantum}
Eitan Geva and Ronnie Kosloff.
\newblock {The quantum heat engine and heat pump: An irreversible thermodynamic
  analysis of the three-level amplifier}.
\newblock \emph{The Journal of chemical physics}, 104\penalty0 (19):\penalty0
  7681--7699, 1996.

\bibitem[Gluza et~al.(2021)Gluza, Sabino, Ng, Vitagliano, Pezzutto, Omar,
  Mazets, Huber, Schmiedmayer, and Eisert]{PRXQuantum.2.030310}
Marek Gluza, Jo\~ao Sabino, Nelly~H.Y. Ng, Giuseppe Vitagliano, Marco Pezzutto,
  Yasser Omar, Igor Mazets, Marcus Huber, J\"org Schmiedmayer, and Jens Eisert.
\newblock {Quantum Field Thermal Machines}.
\newblock \emph{PRX Quantum}, 2:\penalty0 030310, Jul 2021.
\newblock \doi{10.1103/PRXQuantum.2.030310}.
\newblock URL \url{https://link.aps.org/doi/10.1103/PRXQuantum.2.030310}.

\bibitem[Gorini et~al.(1976)Gorini, Kossakowski, and Sudarshan]{1.522979}
Vittorio Gorini, Andrzej Kossakowski, and E.~C.~G. Sudarshan.
\newblock {Completely positive dynamical semigroups of N‐level systems}.
\newblock \emph{Journal of Mathematical Physics}, 17\penalty0 (5):\penalty0
  821--825, 05 1976.
\newblock ISSN 0022-2488.
\newblock \doi{10.1063/1.522979}.
\newblock URL \url{https://doi.org/10.1063/1.522979}.

\bibitem[Hatano(2019)]{Hatano18082019}
Naomichi Hatano.
\newblock {Exceptional points of the Lindblad operator of a two-level system}.
\newblock \emph{Molecular Physics}, 117\penalty0 (15-16):\penalty0 2121--2127,
  2019.
\newblock \doi{10.1080/00268976.2019.1593535}.
\newblock URL \url{https://doi.org/10.1080/00268976.2019.1593535}.

\bibitem[Hatano and Nelson(1997)]{PhysRevB.56.8651}
Naomichi Hatano and David~R. Nelson.
\newblock {{Vortex pinning and non-Hermitian quantum mechanics}}.
\newblock \emph{Phys. Rev. B}, 56:\penalty0 8651--8673, Oct 1997.
\newblock \doi{10.1103/PhysRevB.56.8651}.
\newblock URL \url{https://link.aps.org/doi/10.1103/PhysRevB.56.8651}.

\bibitem[Hatano and Ordonez(2014)]{10.1063/1.4904200}
Naomichi Hatano and Gonzalo Ordonez.
\newblock Time-reversal symmetric resolution of unity without background
  integrals in open quantum systems.
\newblock \emph{Journal of Mathematical Physics}, 55\penalty0 (12):\penalty0
  122106, 12 2014.
\newblock ISSN 0022-2488.
\newblock \doi{10.1063/1.4904200}.
\newblock URL \url{https://doi.org/10.1063/1.4904200}.

\bibitem[Haus and Mullen(1962)]{haus1962quantum}
Hermann~A Haus and JA~Mullen.
\newblock {Quantum noise in linear amplifiers}.
\newblock \emph{Physical Review}, 128\penalty0 (5):\penalty0 2407, 1962.

\bibitem[Heiss(2012)]{Heiss_2012}
W~D Heiss.
\newblock {The physics of exceptional points}.
\newblock \emph{Journal of Physics A: Mathematical and Theoretical},
  45\penalty0 (44):\penalty0 444016, oct 2012.
\newblock \doi{10.1088/1751-8113/45/44/444016}.
\newblock URL \url{https://dx.doi.org/10.1088/1751-8113/45/44/444016}.

\bibitem[Heiss(2004)]{heiss2004exceptional}
WD~Heiss.
\newblock {Exceptional points of non-Hermitian operators}.
\newblock \emph{Journal of Physics A: Mathematical and General}, 37\penalty0
  (6):\penalty0 2455, 2004.

\bibitem[Hewgill et~al.(2021)Hewgill, De~Chiara, and
  Imparato]{PhysRevResearch.3.013165}
Adam Hewgill, Gabriele De~Chiara, and Alberto Imparato.
\newblock {Quantum thermodynamically consistent local master equations}.
\newblock \emph{Phys. Rev. Res.}, 3:\penalty0 013165, Feb 2021.
\newblock \doi{10.1103/PhysRevResearch.3.013165}.
\newblock URL \url{https://link.aps.org/doi/10.1103/PhysRevResearch.3.013165}.

\bibitem[Hofer et~al.(2017)Hofer, Perarnau-Llobet, Miranda, Haack, Silva,
  Brask, and Brunner]{Hofer_2017}
Patrick~P Hofer, Mar\'t Perarnau-Llobet, L~David~M Miranda, \'Graldine Haack,
  Ralph Silva, Jonatan~Bohr Brask, and Nicolas Brunner.
\newblock {Markovian master equations for quantum thermal machines: local
  versus global approach}.
\newblock \emph{New Journal of Physics}, 19\penalty0 (12):\penalty0 123037, dec
  2017.
\newblock \doi{10.1088/1367-2630/aa964f}.
\newblock URL \url{https://dx.doi.org/10.1088/1367-2630/aa964f}.

\bibitem[Ishizaki et~al.(2023)Ishizaki, Hatano, and
  Tajima]{PhysRevResearch.5.023066}
Miku Ishizaki, Naomichi Hatano, and Hiroyasu Tajima.
\newblock Switching the function of the quantum otto cycle in non-markovian
  dynamics: Heat engine, heater, and heat pump.
\newblock \emph{Phys. Rev. Res.}, 5:\penalty0 023066, Apr 2023.
\newblock \doi{10.1103/PhysRevResearch.5.023066}.
\newblock URL \url{https://link.aps.org/doi/10.1103/PhysRevResearch.5.023066}.

\bibitem[Kashuba and Schoeller(2013)]{PhysRevB.87.201402}
Oleksiy Kashuba and Herbert Schoeller.
\newblock {Transient dynamics of open quantum systems}.
\newblock \emph{Phys. Rev. B}, 87:\penalty0 201402, May 2013.
\newblock \doi{10.1103/PhysRevB.87.201402}.
\newblock URL \url{https://link.aps.org/doi/10.1103/PhysRevB.87.201402}.

\bibitem[Khandelwal et~al.(2021)Khandelwal, Brunner, and
  Haack]{PRXQuantum.2.040346}
Shishir Khandelwal, Nicolas Brunner, and G\'eraldine Haack.
\newblock {Signatures of Liouvillian Exceptional Points in a Quantum Thermal
  Machine}.
\newblock \emph{PRX Quantum}, 2:\penalty0 040346, Dec 2021.
\newblock \doi{10.1103/PRXQuantum.2.040346}.
\newblock URL \url{https://link.aps.org/doi/10.1103/PRXQuantum.2.040346}.

\bibitem[Kieu(2004)]{kieu2004second}
Tien~D Kieu.
\newblock {The second law, Maxwell's demon, and work derivable from quantum
  heat engines}.
\newblock \emph{Physical review letters}, 93\penalty0 (14):\penalty0 140403,
  2004.

\bibitem[Kosloff(1984)]{kosloff1984quantum}
Ronnie Kosloff.
\newblock {A quantum mechanical open system as a model of a heat engine}.
\newblock \emph{The Journal of chemical physics}, 80\penalty0 (4):\penalty0
  1625--1631, 1984.

\bibitem[Kosloff et~al.(2000)Kosloff, Geva, and Gordon]{kosloff2000quantum}
Ronnie Kosloff, Eitan Geva, and Jeffrey~M Gordon.
\newblock {Quantum refrigerators in quest of the absolute zero}.
\newblock \emph{Journal of Applied Physics}, 87\penalty0 (11):\penalty0
  8093--8097, 2000.

\bibitem[Lee et~al.(2009)Lee, Yang, Moon, Lee, Shim, Kim, Lee, and
  An]{PhysRevLett.103.134101}
Sang-Bum Lee, Juhee Yang, Songky Moon, Soo-Young Lee, Jeong-Bo Shim, Sang~Wook
  Kim, Jai-Hyung Lee, and Kyungwon An.
\newblock {Observation of an Exceptional Point in a Chaotic Optical
  Microcavity}.
\newblock \emph{Phys. Rev. Lett.}, 103:\penalty0 134101, Sep 2009.
\newblock \doi{10.1103/PhysRevLett.103.134101}.
\newblock URL \url{https://link.aps.org/doi/10.1103/PhysRevLett.103.134101}.

\bibitem[Levy and Kosloff(2012)]{PhysRevLett.108.070604}
Amikam Levy and Ronnie Kosloff.
\newblock {Quantum Absorption Refrigerator}.
\newblock \emph{Phys. Rev. Lett.}, 108:\penalty0 070604, Feb 2012.
\newblock \doi{10.1103/PhysRevLett.108.070604}.
\newblock URL \url{https://link.aps.org/doi/10.1103/PhysRevLett.108.070604}.

\bibitem[Liao et~al.(2021)Liao, Leblanc, Ren, Li, Li, Solnyshkov, Malpuech,
  Yao, and Fu]{PhysRevLett.127.107402}
Qing Liao, Charly Leblanc, Jiahuan Ren, Feng Li, Yiming Li, Dmitry Solnyshkov,
  Guillaume Malpuech, Jiannian Yao, and Hongbing Fu.
\newblock {Experimental Measurement of the Divergent Quantum Metric of an
  Exceptional Point}.
\newblock \emph{Phys. Rev. Lett.}, 127:\penalty0 107402, Sep 2021.
\newblock \doi{10.1103/PhysRevLett.127.107402}.
\newblock URL \url{https://link.aps.org/doi/10.1103/PhysRevLett.127.107402}.

\bibitem[Lindblad(1976)]{BF01608499}
G.~Lindblad.
\newblock {On the generators of quantum dynamical semigroups}.
\newblock \emph{Communications in Mathematical Physics}, 48\penalty0
  (2):\penalty0 119--130, 1976.
\newblock \doi{10.1007/BF01608499}.
\newblock URL \url{https://doi.org/10.1007/BF01608499}.

\bibitem[Lloyd(1997)]{lloyd1997quantum}
Seth Lloyd.
\newblock {Quantum-mechanical Maxwell’s demon}.
\newblock \emph{Physical Review A}, 56\penalty0 (5):\penalty0 3374, 1997.

\bibitem[Loudon and Shepherd(1984)]{loudon1984properties}
Rodney Loudon and TJ~Shepherd.
\newblock {Properties of the optical quantum amplifier}.
\newblock \emph{Optica Acta: International Journal of Optics}, 31\penalty0
  (11):\penalty0 1243--1269, 1984.

\bibitem[Manikandan et~al.(2020)Manikandan, Jussiau, and
  Jordan]{PhysRevB.102.235427}
Sreenath~K. Manikandan, \'Etienne Jussiau, and Andrew~N. Jordan.
\newblock {Autonomous quantum absorption refrigerators}.
\newblock \emph{Phys. Rev. B}, 102:\penalty0 235427, Dec 2020.
\newblock \doi{10.1103/PhysRevB.102.235427}.
\newblock URL \url{https://link.aps.org/doi/10.1103/PhysRevB.102.235427}.

\bibitem[Manzano(2020)]{10.1063/1.5115323}
Daniel Manzano.
\newblock {A short introduction to the Lindblad master equation}.
\newblock \emph{AIP Advances}, 10\penalty0 (2):\penalty0 025106, 02 2020.
\newblock ISSN 2158-3226.
\newblock \doi{10.1063/1.5115323}.
\newblock URL \url{https://doi.org/10.1063/1.5115323}.

\bibitem[Maslennikov et~al.(2019)Maslennikov, Ding, Habl{\"u}tzel, Gan, Roulet,
  Nimmrichter, Dai, Scarani, and Matsukevich]{Maslennikov2019}
Gleb Maslennikov, Shiqian Ding, Roland Habl{\"u}tzel, Jaren Gan, Alexandre
  Roulet, Stefan Nimmrichter, Jibo Dai, Valerio Scarani, and Dzmitry
  Matsukevich.
\newblock {Quantum absorption refrigerator with trapped ions}.
\newblock \emph{Nature Communications}, 10\penalty0 (1):\penalty0 202, 2019.
\newblock \doi{10.1038/s41467-018-08090-0}.
\newblock URL \url{https://doi.org/10.1038/s41467-018-08090-0}.

\bibitem[Minganti et~al.(2019)Minganti, Miranowicz, Chhajlany, and
  Nori]{PhysRevA.100.062131}
Fabrizio Minganti, Adam Miranowicz, Ravindra~W. Chhajlany, and Franco Nori.
\newblock {Quantum exceptional points of non-Hermitian Hamiltonians and
  Liouvillians: The effects of quantum jumps}.
\newblock \emph{Phys. Rev. A}, 100:\penalty0 062131, Dec 2019.
\newblock \doi{10.1103/PhysRevA.100.062131}.
\newblock URL \url{https://link.aps.org/doi/10.1103/PhysRevA.100.062131}.

\bibitem[Miri and Al{\`u}(2019)]{doi:10.1126/science.aar7709}
Mohammad-Ali Miri and Andrea Al{\`u}.
\newblock {Exceptional points in optics and photonics}.
\newblock \emph{Science}, 363\penalty0 (6422):\penalty0 eaar7709, 2019.
\newblock \doi{10.1126/science.aar7709}.
\newblock URL \url{https://www.science.org/doi/abs/10.1126/science.aar7709}.

\bibitem[Moiseyev(2011)]{moiseyev2011non}
Nimrod Moiseyev.
\newblock \emph{{Non-Hermitian quantum mechanics}}.
\newblock Cambridge University Press, 2011.

\bibitem[Nakajima(1958)]{10.1143/PTP.20.948}
Sadao Nakajima.
\newblock {On Quantum Theory of Transport Phenomena: Steady Diffusion}.
\newblock \emph{Progress of Theoretical Physics}, 20\penalty0 (6):\penalty0
  948--959, 12 1958.
\newblock ISSN 0033-068X.
\newblock \doi{10.1143/PTP.20.948}.
\newblock URL \url{https://doi.org/10.1143/PTP.20.948}.

\bibitem[Palao et~al.(2001)Palao, Kosloff, and Gordon]{palao2001quantum}
Jos{\'e}~P Palao, Ronnie Kosloff, and Jeffrey~M Gordon.
\newblock {Quantum thermodynamic cooling cycle}.
\newblock \emph{Physical Review E}, 64\penalty0 (5):\penalty0 056130, 2001.

\bibitem[Partovi(1989)]{partovi1989quantum}
M~Hossein Partovi.
\newblock {Quantum thermodynamics}.
\newblock \emph{Physics Letters A}, 137\penalty0 (9):\penalty0 440--444, 1989.

\bibitem[Quan et~al.(2007)Quan, Liu, Sun, and Nori]{quan2007quantum}
Hai-Tao Quan, Yu-xi Liu, Chang-Pu Sun, and Franco Nori.
\newblock {Quantum thermodynamic cycles and quantum heat engines}.
\newblock \emph{Physical Review E—Statistical, Nonlinear, and Soft Matter
  Physics}, 76\penalty0 (3):\penalty0 031105, 2007.

\bibitem[Rahmani et~al.(2024)Rahmani, Opala, and
  Matuszewski]{PhysRevB.109.085311}
Amir Rahmani, Andrzej Opala, and Micha\l{} Matuszewski.
\newblock {Exceptional points and phase transitions in non-Hermitian nonlinear
  binary systems}.
\newblock \emph{Phys. Rev. B}, 109:\penalty0 085311, Feb 2024.
\newblock \doi{10.1103/PhysRevB.109.085311}.
\newblock URL \url{https://link.aps.org/doi/10.1103/PhysRevB.109.085311}.

\bibitem[Roccati et~al.(2022)Roccati, Palma, Ciccarello, and
  Bagarello]{doi:10.1142/S1230161222500044}
Federico Roccati, G.~Massimo Palma, Francesco Ciccarello, and Fabio Bagarello.
\newblock {Non-Hermitian Physics and Master Equations}.
\newblock \emph{Open Systems \& Information Dynamics}, 29\penalty0
  (01):\penalty0 2250004, 2022.
\newblock \doi{10.1142/S1230161222500044}.
\newblock URL \url{https://doi.org/10.1142/S1230161222500044}.

\bibitem[Rotter(2009)]{rotter2009non}
Ingrid Rotter.
\newblock {A non-Hermitian Hamilton operator and the physics of open quantum
  systems}.
\newblock \emph{Journal of Physics A: Mathematical and Theoretical},
  42\penalty0 (15):\penalty0 153001, 2009.

\bibitem[Santos and Chattopadhyay(2023)]{santos2023pt}
Jonas~FG Santos and Pritam Chattopadhyay.
\newblock {PT-symmetry effects in measurement-based quantum thermal machines}.
\newblock \emph{Physica A: Statistical Mechanics and its Applications},
  632:\penalty0 129342, 2023.

\bibitem[Schaller(2014)]{schaller2014open}
G.~Schaller.
\newblock \emph{{Open Quantum Systems Far from Equilibrium}}.
\newblock Lecture Notes in Physics. Springer International Publishing, 2014.
\newblock ISBN 9783319038773.
\newblock URL \url{https://books.google.co.jp/books?id=deS5BQAAQBAJ}.

\bibitem[Scovil and Schulz-DuBois(1959)]{PhysRevLett.2.262}
H.~E.~D. Scovil and E.~O. Schulz-DuBois.
\newblock {Three-Level Masers as Heat Engines}.
\newblock \emph{Phys. Rev. Lett.}, 2:\penalty0 262--263, Mar 1959.
\newblock \doi{10.1103/PhysRevLett.2.262}.
\newblock URL \url{https://link.aps.org/doi/10.1103/PhysRevLett.2.262}.

\bibitem[Segal(2018)]{PhysRevE.97.052145}
Dvira Segal.
\newblock {Current fluctuations in quantum absorption refrigerators}.
\newblock \emph{Phys. Rev. E}, 97:\penalty0 052145, May 2018.
\newblock \doi{10.1103/PhysRevE.97.052145}.
\newblock URL \url{https://link.aps.org/doi/10.1103/PhysRevE.97.052145}.

\bibitem[Shibata et~al.(1977)Shibata, Takahashi, and Hashitsume]{BF01040100}
Fumiaki Shibata, Yoshinori Takahashi, and Natsuki Hashitsume.
\newblock {A generalized stochastic liouville equation. Non-Markovian versus
  memoryless master equations}.
\newblock \emph{Journal of Statistical Physics}, 17\penalty0 (4):\penalty0
  171--187, 1977.
\newblock \doi{10.1007/BF01040100}.
\newblock URL \url{https://doi.org/10.1007/BF01040100}.

\bibitem[Shirai et~al.(2021)Shirai, Hashimoto, Tezuka, Uchiyama, and
  Hatano]{PhysRevResearch.3.023078}
Yuji Shirai, Kazunari Hashimoto, Ryuta Tezuka, Chikako Uchiyama, and Naomichi
  Hatano.
\newblock Non-markovian effect on quantum otto engine: Role of system-reservoir
  interaction.
\newblock \emph{Phys. Rev. Res.}, 3:\penalty0 023078, Apr 2021.
\newblock \doi{10.1103/PhysRevResearch.3.023078}.
\newblock URL \url{https://link.aps.org/doi/10.1103/PhysRevResearch.3.023078}.

\bibitem[Silva et~al.(2015)Silva, Skrzypczyk, and Brunner]{PhysRevE.92.012136}
Ralph Silva, Paul Skrzypczyk, and Nicolas Brunner.
\newblock {Small quantum absorption refrigerator with reversed couplings}.
\newblock \emph{Phys. Rev. E}, 92:\penalty0 012136, Jul 2015.
\newblock \doi{10.1103/PhysRevE.92.012136}.
\newblock URL \url{https://link.aps.org/doi/10.1103/PhysRevE.92.012136}.

\bibitem[Singh et~al.(2020)Singh, Pandit, and Johal]{PhysRevE.101.062121}
Varinder Singh, Tanmoy Pandit, and Ramandeep~S. Johal.
\newblock {Optimal performance of a three-level quantum refrigerator}.
\newblock \emph{Phys. Rev. E}, 101:\penalty0 062121, Jun 2020.
\newblock \doi{10.1103/PhysRevE.101.062121}.
\newblock URL \url{https://link.aps.org/doi/10.1103/PhysRevE.101.062121}.

\bibitem[Stenholm(1986)]{stenholm1986theory}
Stig Stenholm.
\newblock {The theory of quantum amplifiers}.
\newblock \emph{Physica Scripta}, 1986\penalty0 (T12):\penalty0 56, 1986.

\bibitem[Zhang et~al.(2022)Zhang, Zhang, Ding, Li, Bu, Wang, Yan, Su, Chen,
  Nori, et~al.]{zhang2022dynamical}
J-W Zhang, J-Q Zhang, G-Y Ding, J-C Li, J-T Bu, B~Wang, L-L Yan, S-L Su,
  L~Chen, F~Nori, et~al.
\newblock {Dynamical control of quantum heat engines using exceptional points}.
\newblock \emph{Nature Communications}, 13\penalty0 (1):\penalty0 6225, 2022.

\bibitem[Zwanzig(1960)]{10.1063/1.1731409}
Robert Zwanzig.
\newblock {Ensemble Method in the Theory of Irreversibility}.
\newblock \emph{The Journal of Chemical Physics}, 33\penalty0 (5):\penalty0
  1338--1341, 11 1960.
\newblock ISSN 0021-9606.
\newblock \doi{10.1063/1.1731409}.
\newblock URL \url{https://doi.org/10.1063/1.1731409}.

\end{thebibliography}
\end{document}